\documentclass[a4paper,11pt]{article}
\pdfoutput=1 

\usepackage{jcappub}
\usepackage{graphicx}
\usepackage{dcolumn}
\usepackage{amssymb,amsmath,bm}
\usepackage{color}
\usepackage[dvipsnames]{xcolor}
\usepackage{xfrac}
\usepackage{aas_macros}
\usepackage{mathrsfs}
\usepackage{subcaption}
\usepackage{rotating}
\usepackage{chngcntr}
\usepackage[T1]{fontenc}
\usepackage{natbib}
\usepackage{lipsum}
\usepackage{xspace}
\usepackage{lineno}
\usepackage{multirow}
\usepackage{soul}
\setstcolor{red}

\newcommand{\bk}{BICEP2/{\sl Keck Array}~}
\newcommand{\planck}{{\sl Planck}~}
\newcommand{\wmap}{{\sl WMAP}~}
\newcommand{\fsig}{$f_{\mbox{\scriptsize sig}}$\xspace}
\newcommand{\frot}{$f_{\mbox{\scriptsize rot}}$\xspace}
\newcommand{\fdb}{$f_{\mbox{\scriptsize 3dB}}$\xspace}

\title{The Simons Observatory: gain, bandpass and polarization-angle calibration requirements for B-mode searches}

\author[1]{Maximilian H. Abitbol,}
\author[1]{David Alonso,}
\author[2]{Sara M. Simon,}
\author[3]{Jack Lashner,}
\author[4]{Kevin T. Crowley,}

\author[4]{Aamir M. Ali,}
\author[1]{Susanna Azzoni,}
\author[5,6,7]{Carlo Baccigalupi,}
\author[30]{Darcy Barron,}
\author[12]{Michael L. Brown,}
\author[8]{Erminia Calabrese,}
\author[9]{Julien Carron,}
\author[10,11]{Yuji Chinone,}
\author[12]{Jens Chluba,} 
\author[13]{Gabriele Coppi,}
\author[16]{Kevin D. Crowley,}
\author[14]{Mark Devlin,}
\author[15,16]{Jo Dunkley,}
\author[17]{Josquin Errard,}
\author[13]{Valentina Fanfani,}
\author[18]{Nicholas Galitzki,}
\author[31]{Martina Gerbino,}
\author[32,33]{J. Colin Hill,}
\author[19]{Bradley R. Johnson,}
\author[17]{Baptiste Jost,}
\author[18]{Brian Keating,}
\author[5]{Nicoletta Krachmalnicoff,}
\author[10,27,36,37]{Akito Kusaka,}
\author[20]{Adrian T. Lee,}
\author[21]{Thibaut Louis,}
\author[38]{Mathew S. Madhavacheril,}
\author[16]{Heather McCarrick,}
\author[22,23,24,25]{Jeffrey McMahon,}
\author[26]{P. Daniel Meerburg,}
\author[13]{Federico Nati,}
\author[10]{Haruki Nishino,}
\author[16]{Lyman A. Page,}
\author[5,6,7]{Davide Poletti,}
\author[34,35]{Giuseppe Puglisi}
\author[18]{Michael J. Randall,}
\author[12]{Aditya Rotti,}
\author[18]{Jacob Spisak,}
\author[27]{Aritoki Suzuki,}
\author[18]{Grant P. Teply,}
\author[17]{Clara Verg\`es,}
\author[28]{Edward J. Wollack,}
\author[14,29]{Zhilei Xu,}
\author[13]{Mario Zannoni}

\affiliation[1]{University of Oxford, Department of Physics, Denys Wilkinson Building, Keble Road, Oxford OX1 4LS, UK}
\affiliation[2]{Fermi National Accelerator Laboratory, Batavia, IL, USA}
\affiliation[3]{University of Southern California, Los Angeles, CA, USA}
\affiliation[4]{University of California Berkeley, Berkeley, CA, USA}
\affiliation[5]{International School for Advanced Studies (SISSA), Via Bonomea 265, 34136 Trieste, Italy}
\affiliation[6]{Institute for Fundamental Physics of the Universe (IFPU), Via Beirut 2, 34014 Trieste, Italy}
\affiliation[7]{National Institute for Nuclear Physics (INFN), via Valerio 2, 34127 Trieste, Italy}
\affiliation[8]{School of Physics and Astronomy, Cardiff University, The Parade, CF243AA, Cardiff, UK}
\affiliation[9]{Universit\'e de Gen\`eve, D\'epartement de Physique Th\'eorique et CAP, 24 Quai Ansermet, CH-1211 Gen\`eve 4, Switzerland}
\affiliation[10]{Research Center for the Early Universe, School of Science, The University of Tokyo, Tokyo 113-0033, Japan}
\affiliation[11]{Kavli Institute for the Physics and Mathematics of the Universe (Kavli IPMU, WPI), UTIAS, The University of Tokyo, Kashiwa, Chiba 277-8583, Japan}
\affiliation[12]{Jodrell Bank Centre for Astrophysics, School of Physics and Astronomy, University of Manchester, Oxford Road, Manchester M13 9PL, UK}
\affiliation[13]{Department of Physics, University of Milano - Bicocca, Piazza della Scienza, 3 - 20126 Milano (MI), Italy}
\affiliation[14]{Department of Physics and Astronomy, University of Pennsylvania, 209 South 33rd Street, Philadelphia, PA 19104, USA}
\affiliation[15]{Department of Astrophysical Sciences, Peyton Hall, Princeton University,  Princeton, NJ08544, USA}
\affiliation[16]{Joseph Henry Laboratories of Physics, Jadwin Hall, Princeton University,  Princeton, NJ 08544, USA}
\affiliation[17]{Universit\'e de Paris, CNRS, Astroparticule et Cosmologie, F-75006 Paris, France}
\affiliation[18]{University of California San Diego, 9500 Gilman Dr., La Jolla, CA 92093, USA}
\affiliation[19]{University of Virginia, Department of Astronomy, Charlottesville, VA 22904, USA}
\affiliation[20]{Department of Physics, University of California, Berkeley, CA 94720, USA}
\affiliation[21]{Universit\'e Paris-Saclay, CNRS/IN2P3, IJCLab, 91405 Orsay, France}
\affiliation[22]{Kavli Institute for Cosmological Physics, University of Chicago, 5640 S. Ellis Ave., Chicago, IL 60637, USA}
\affiliation[23]{Department of Astronomy and Astrophysics, University of Chicago, 5640 S. Ellis Ave., Chicago, IL 60637, USA}
\affiliation[24]{Department of Physics, University of Chicago, Chicago, IL 60637, USA}
\affiliation[25]{Enrico Fermi Institute, University of Chicago, Chicago, IL 60637, USA}
\affiliation[26]{Van Swinderen Institute for Particle Physics and Gravity, University of Groningen, Nijenborgh 4, 9747 AG Groningen, The Netherlands}
\affiliation[27]{Physics Division, Lawrence Berkeley National Laboratory, 1 Cyclotron Road, Berkeley, CA 94720, US}
\affiliation[28]{NASA / Goddard Space Flight Center, Greenbelt, MD, 20771, USA}
\affiliation[29]{MIT Kavli Institute, Massachusetts Institute of Technology, 77 Massachusetts Avenue, Cambridge, MA 02139, USA}
\affiliation[30]{Department of Physics and Astronomy, University of New Mexico, Albuquerque, NM 87131, USA}
\affiliation[31]{Istituto Nazionale di Fisica Nucleare, sezione di Ferrara,
Polo Scientifico e Tecnologico - Edificio C Via Saragat, 1, I-44122, Ferrara, Italy}
\affiliation[32]{Center for Computational Astrophysics, Flatiron Institute, 162 5th Avenue, New York, NY, USA 10010}
\affiliation[33]{Department of Physics, Columbia University, 538 West 120th Street, New York, NY, USA 10027}
\affiliation[34]{Space Sciences Laboratory,  University of California, Berkeley, CA 94720,  USA}
\affiliation[35]{Computational Cosmology Center, Lawrence Berkeley National Laboratory, Berkeley, CA 94720, USA}
\affiliation[36]{Department of Physics, The University of Tokyo, Tokyo 113-0033, Japan}
\affiliation[37]{Kavli Institute for the Physics and Mathematics of the Universe (WPI), Berkeley Satellite, the University of California, Berkeley 94720, USA}
\affiliation[38]{Perimeter Institute for Theoretical Physics, 31 Caroline Street N, Waterloo ON N2L 2Y5 Canada}

\emailAdd{maximilian.abitbol@physics.ox.ac.uk}

\abstract{We quantify the calibration requirements for systematic uncertainties for next-generation ground-based observatories targeting the large-angle $B$-mode polarization of the Cosmic Microwave Background, with a focus on the Simons Observatory (SO). We explore uncertainties on gain calibration, bandpass center frequencies, and polarization angles, including the frequency variation of the latter across the bandpass. We find that gain calibration and bandpass center frequencies must be known to percent levels or less to avoid biases on the tensor-to-scalar ratio $r$ on the order of $\Delta r\sim10^{-3}$, in line with previous findings. Polarization angles must be calibrated to the level of a few tenths of a degree, while their frequency variation between the edges of the band must be known to ${\cal O}(10)$ degrees. Given the tightness of these calibration requirements, we explore the level to which residual uncertainties on these systematics would affect the final constraints on $r$ if included in the data model and marginalized over. We find that the additional parameter freedom does not degrade the final constraints on $r$ significantly, broadening the error bar by ${\cal O}(10\%)$ at most. We validate these results by reanalyzing the latest publicly available data from the \bk collaboration within an extended parameter space covering both cosmological, foreground and systematic parameters. Finally, our results are discussed in light of the instrument design and calibration studies carried out within SO.}

\begin{document}
\maketitle
\flushbottom

\section{Introduction}\label{sec:intro}
The search for primordial $B$-mode polarization is one of the primary endeavours for understanding the physics of the early Universe with the Cosmic Microwave Background (CMB). If associated with primordial tensor perturbations \citep{1997PhRvL..78.2058K, 1997PhRvL..78.2054S}, their detection with a sufficiently large amplitude would rule out large classes of inflationary and non-inflationary models. The primordial $B$-mode amplitude is usually parametrized in terms of the tensor-to-scalar ratio $r$. Although inflation is able to generate tensor fluctuation with an arbitrarily small amplitude, a large family of models, such as Higgs or $R^2$ inflation \cite{1979JETPL..30..682S,2008PhLB..659..703B}, predict values for $r\propto1/N^2\sim0.001$, where $N$ is the number of $e$-folds of inflation. These models are particularly interesting, given their connection with Standard Model physics or quantum corrections to gravity. Current constraints on $r$ from the BICEP2/{\it Keck} are at the level of $r<0.07$ \cite{2016PhRvL.116c1302B} (or $r<0.044$ in combination with \planck{} \citep{2020arXiv201001139T}). Improving on these limits is the goal of ongoing experiments, such as CLASS, POLARBEAR, and the South Pole Telescope \cite{2014SPIE.9153E..1IE,2019arXiv191002608A,2014SPIE.9153E..1PB,2014JLTP..176..733M}. Next-generation observatories that will come online in the 2020s, including the Simons Observatory, BICEP Array, LiteBIRD, and CMB Stage-4, have been designed to reach sensitivities that would enable a  statistical uncertainty on the tensor-to-scalar ratio at the level of $\sigma(r)\approx10^{-3}$ or lower \cite{2016arXiv161002743A,2018SPIE10708E..07H,2018PhRvL.121v1301B,2018SPIE10698E..4FS,2019JCAP...02..056A}. The actual achieved sensitivity, however, will depend on the impact of instrumental and astrophysical systematics.

$B$-modes, the parity-odd component of the CMB polarization field, are sourced by primordial tensor fluctuations within the standard inflationary paradigm, while scalar perturbations are only able to generate $E$-mode polarization. The faintness of the primordial $B$-mode signal, however, makes it sub-dominant compared to other sources of $B$-modes, including Galactic foregrounds and those caused by the lensing of the CMB $E$-modes. Therefore, disentangling the different sky components in a reliable way is the most significant analysis challenge for these experiments \cite{2014PhRvL.112x1101B,2015PhRvL.114j1301B,2016MNRAS.458.2032R,2016JCAP...03..052E,2017PhRvD..95d3504A,2017MNRAS.468.4408H,2019arXiv191002608A}.

The faintness of the signal imposes tight constraints on the instrument design and calibration requirements. Component separation disentangles foregrounds and the CMB through their different frequency dependence, which relies heavily on understanding the frequency dependence of the instrument itself. This includes the frequency transmission curves for each band, as well as the frequency dependence of the most relevant instrumental effects, such as polarization angles or beams. Since the primordial $B$-mode signal peaks on large angular scales, with the so-called recombination bump at $\ell\sim80$, ground-based observatories frequently employ modulation schemes to access large angular scales, for example through the use of cryogenic half-wave plates (HWPs) to reduce the impact of atmospheric noise, and to mitigate certain systematics from, for example, pair-differencing \citep{kusakaModulationCMBPolarization2014,essinger-hilemanSystematicEffectsAmbienttemperature2016,2017JCAP...05..008T,ABS2018}, as well as other instrumental sources of $1/f$ noise. The systematic effects associated with HWPs must also be thoroughly understood.

The Simons Observatory (SO \cite{2019JCAP...02..056A}), with its Small Aperture Telescopes (SATs), will produce high-sensitivity maps of the CMB polarization at the degree-scale across 10\% of the sky ($\sim4,000\,{\rm deg}^2$) across a broad range of frequencies ($27-280\,{\rm GHz}$)~\cite{ali2020}. In order to achieve its goal of constraining $r$ at the level of $\sigma(r)\simeq0.002$\footnote{Note that although this sensitivity will not allow a detection of a $r\sim0.001$ signal, it will constitute an order of magnitude improvement over current bounds, and will allow the detection of an $r\sim0.01$ signal. This is not a fundamental limit, however, and could be improved by extending the number of detectors or the observing time.}, SO will need to tackle both instrument and analysis challenges. This paper presents a comprehensive study of the calibration requirements associated with bandpass and polarization angle systematics in SO. Using a multi-frequency power-spectrum-based analysis pipeline, we try to answer two questions for an SO-like dataset:
\begin{enumerate}
    \item To what level do these systematics need to be calibrated so as not to significantly bias the final constraints on $r$ if the residual systematic effects are ignored?
    \item How much do the final $r$ constraints degrade if the residual systematics are modeled and marginalized over?
\end{enumerate}
The results found through our analysis are (summarized in Table~\ref{tab:conclusion}):
\begin{enumerate}
    \item Gains and bandpass central frequencies must be calibrated to the $\lesssim 0.5\%$ level, polarization angles must be calibrated to $\lesssim 0.2^\circ$, and their frequency variation within the band must be known to within $\sim 10^\circ$ (see Table \ref{tab:syst_req_table}).
    \item Assuming a $\mathcal{O}(3\%)$ uncertainty on the gains and central frequencies, and nulling the CMB $EB$ cross-spectrum to self-calibrate the polarization angles, we find that the final constraints on $r$ degrade by at most $10\%$ (see Table \ref{tab:marg}). Thus, the requirements above can be relaxed significantly provided a reliable model for the residual systematic effects.
\end{enumerate}

The paper is structured as follows. Section \ref{sec:methods} presents the methods used for our analysis, including the model used to describe the sky signal, the instrument, and the different systematic uncertainties studied here, as well as the methodology to quantify the calibration requirements and their impact on the final $B$-mode constraints. Section \ref{sec:results} presents our results, including both the raw systematic calibration requirements for unmodeled effects as well as the impact of marginalized nuisance parameters on the $r$ constraints. We also present an application of our methods to real CMB polarization data from the \bk collaboration in Section~\ref{sec:bicep}. Section \ref{sec:design} then discusses these results in the context of the SO design and calibration efforts. We present our conclusions in Section \ref{sec:conclusion}. 

\section{Methods}
\label{sec:methods}

\subsection{$B$-mode analysis pipeline}
We use a multi-frequency power spectrum-based component separation method based on that used by \cite{2016PhRvL.116c1302B,2018PhRvL.121v1301B}. In this case, the data vector is the full set of cross-correlations between all frequency maps, $C_\ell^{\alpha\beta}$, and the map-level sky and instrument model, described in the next sections, is propagated to these power spectra.

On large scales, where only a small number of modes are available, the central limit theorem cannot be invoked, and approximating the likelihood of the measured power spectra as Gaussian becomes inaccurate. To account for this effect, we use the non-Gaussian likelihood proposed by \cite{2008PhRvD..77j3013H} (HL here). This requires the use of a fiducial covariance matrix for $C_\ell^{\nu\nu'}$, which we approximate using the so-called \emph{Knox formula} \cite{1997ApJ...480...72K}:
\begin{equation}\label{eq:cov_knox}
{\rm Cov}\left[C^{\alpha\beta}_\ell,C^{\gamma\delta}_{\ell'}\right]=\delta_{\ell\ell'}\frac{C^{\alpha\gamma}_\ell C^{\beta\delta}_\ell+C^{\alpha\delta}_\ell C^{\beta\gamma}_\ell}{(2\ell+1)\,\Delta\ell\,f_{\rm sky}},
\end{equation}
where $\Delta\ell$ is the number of multipoles in the bandpower labelled with multipole $\ell$, and $f_{\rm sky}=0.1$ is the usable sky fraction of the SO $B$-mode footprint \cite{2019JCAP...02..056A}. This approximation is not accurate enough for actual data analysis, which require extensive simulations, but it suffices for the forecasting exercise we carry out here, as demonstrated in \cite{2019JCAP...02..056A}. The power spectra entering Eq. \ref{eq:cov_knox} contain both signal and noise contributions.  All our constraints on $r$ use scale cuts $30\leq \ell\leq 300$, where the lower bound is motivated by the expectation that atmospheric noise and ground pickup will dominate the largest scales.

The HL likelihood also requires an estimate of the fiducial power spectra, including the corresponding noise power spectrum (which should also be included in the power spectra in Eq. \ref{eq:cov_knox}). For this, we use the projected noise curves for the SO Small Aperture Telescope (made publicly available in \cite{2019JCAP...02..056A}\footnote{\url{https://github.com/simonsobs/so_noise_models}}). We use the best-case forecasted noise levels, the \emph{goal} noise level and the \emph{optimistic} knee angular scale $\ell_{\rm knee}$, below which atmospheric noise dominates the spectrum, ranging from $\ell_{\rm knee}=15$ to $\ell_{\rm knee}=40$ across the full frequency range \citep{2019JCAP...02..056A}. The goal noise levels and optimistic $\ell_{\rm knee}$ are used in this work since this results in conservative estimates for the instrument requirements. These estimates, however, may not hold for future, more sensitive experiments, such as CMB Stage-4 \citep{2016arXiv161002743A}.

The posterior parameter distribution is given by the product of the HL likelihood and a set of priors on all foreground and systematic parameters. We explored a variety of prior distributions and found consistent results throughout. In general, we use wide top-hat priors on the CMB and foreground parameters such that the data provides all the constraining power on parameters. The priors on systematic parameters were chosen to correspond to previously achieved numbers.

\begin{figure}[tbp]
\centering
\includegraphics[width=\textwidth]{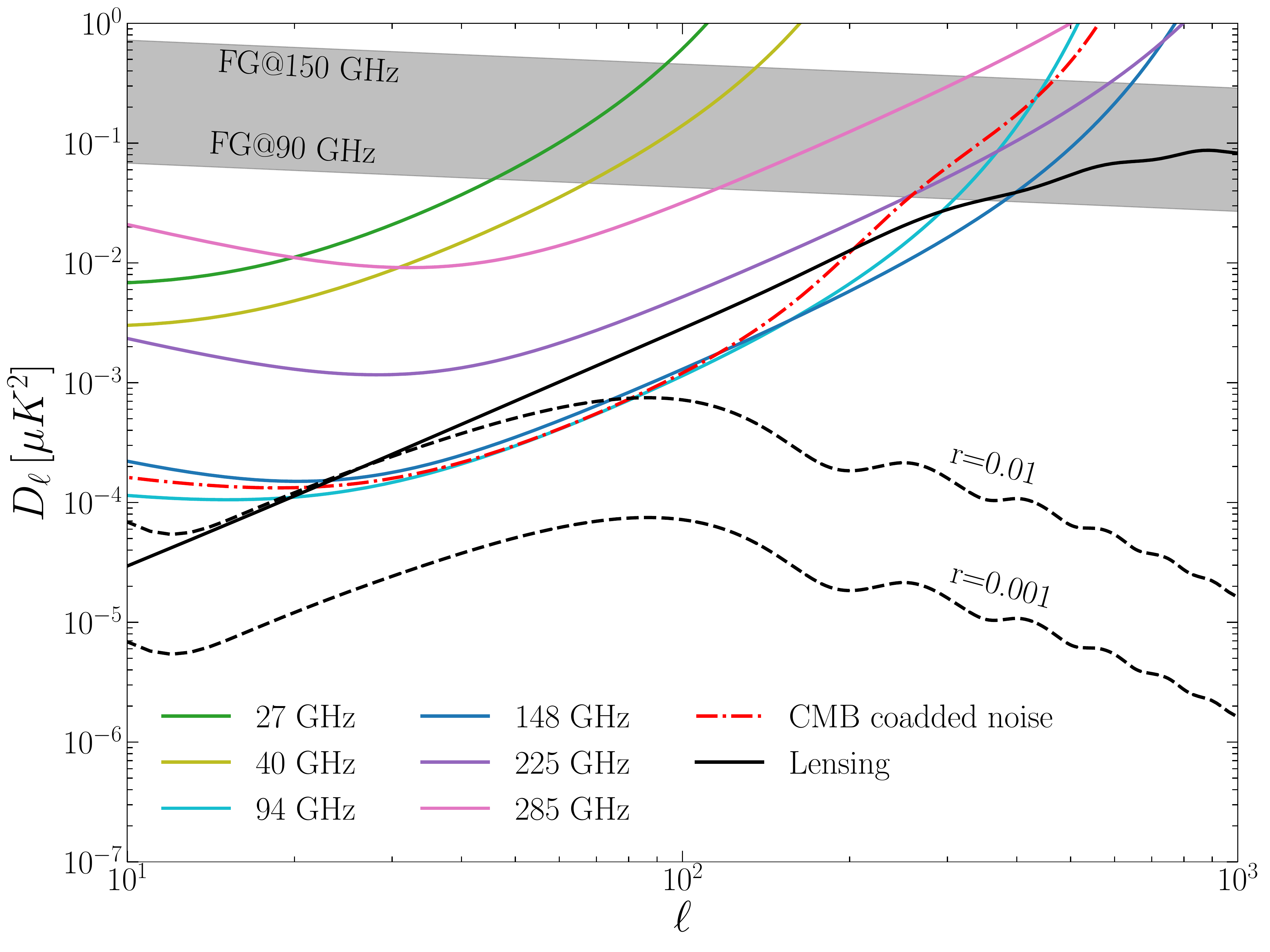}
\caption{Relevant sky signal and noise power spectra. CMB lensing $B$-mode power spectrum (solid black) and contribution from primordial tensor fluctuations for $r=0.01$ and $r=0.001$ (dashed black). The different colored lines show the SO SAT noise power spectra for the six different frequency bands. The gray band shows the estimated combined $B$-mode foreground power spectra between $90\,{\rm GHz}$ and $150\,{\rm GHz}$ for a 10\% sky area. The red dash-dotted line shows the CMB coadded noise after component separation.} \label{fig:noiseps}
\end{figure}

The next two sections describe the models used to describe the different sky components and the instrumental systematic effects, and list the fiducial parameter values used in our calculations. We generate a set of power spectra using those fiducial parameters and use them as a data vector when exploring the posterior parameter distribution. The different contributions to our sky and instrument model are described in Figure~\ref{fig:noiseps}. 

When simply maximizing the posterior we use Powell's minimization method \cite{powell} as implemented in {\tt scipy}\footnote{\url{https://www.scipy.org/}} \cite{2020SciPy-NMeth}. To explore the full posterior distribution we use a (affine invariant ensemble sampler) Markov Chain Monte Carlo approach, as implemented in the {\tt emcee} package\footnote{\url{https://emcee.readthedocs.io}} \cite{2013PASP..125..306F}. We used the chain autocorrelation length, Gelman-Rubin statistic, and manual inspection of the chains to ensure convergence. All chains have at least 10,000 times the autocorrelation length in samples. The Gelman-Rubin statistic is enforced to be less than 1.1, and is generally much smaller than that (<1.01).  We estimate the chain uncertainty by slicing the chains into sub-chains and calculating the standard deviation across sub-chains of the parameter standard deviations within each sub-chain, scaled by the inverse square root of the number of samples in each sub-chain. We find the chain standard deviation of parameter uncertainties to be around 4\% for most cases and less than 10\% for all cases, depending on the chain length and model dimension. 

In all cases we use a smooth data vector with no statistical noise as input to the likelihood for simplicity. Our analysis of the \bk data in Section \ref{sec:bicep} suggests that this does not have a significant impact on our results.

\subsection{Sky model}
\label{ssec:smodel}

At the map level, our sky model is made up of three components: CMB (labelled $C$ here), Galactic synchrotron emission ($S$) and Galactic thermal dust emission ($D$). In the most general case, we consider the cross-correlation between $E$ and $B$ modes between different frequency bands. A single map is therefore labelled by a polarization channel $p\in\{E,\,B\}$, and either a component index $c\in\{C,S,D\}$ or a frequency $\nu$. We will label the cross-correlation between two such maps (labelled by $(p,c)$ and $(p',c')$) as $C^{cc',pp'}_\ell$ here.

Our sky model is summarized at the power spectrum level as
\begin{equation}\label{eq:cl_sky_full}
C^{\nu\nu',pp'}_\ell=f^\nu_Cf^{\nu'}_C\,C_\ell^{CC,pp'}+\left.C^{\nu\nu',pp'}_\ell\right|_{\rm FG},
\end{equation}
where $f^\nu_C$ is the conversion from the CMB spectrum in thermodynamic units to antenna temperature units
\begin{equation}
f^\nu_C=e^x\left(\frac{x}{e^x-1}\right)^2, \hspace{12pt} x\equiv\frac{h\nu}{k_B\,T_{\rm CMB}},
\end{equation}
and where $h$ is Planck's constant, $k_B$ is Boltzmann's constant, and $T_{\rm CMB}=2.7255\,{\rm K}$ is the CMB temperature~\cite{fixsen2009}. $C_\ell^{CC,EE(BB)}$ is the CMB $E$($B$)-mode power spectrum. We parametrize $C_\ell^{CC,BB}$ as
\begin{equation}\label{eq:cl_cmb_bb}
C^{CC,BB}_\ell=A_{\rm lens}\,C^{{\rm lens},BB}_\ell+r\,C^{{\rm tens},BB}_\ell,
\end{equation}
where $C^{{\rm lens},BB}_\ell$ and $C^{{\rm tens},BB}_\ell$ is the $B$-mode power spectrum due to gravitational lensing and primordial tensor fluctuations, the latter for a tensor-to-scalar ratio $r=1$ and a tensor spectral index\footnote{We do not assume a consistency relation between $r$ and $n_t$.} $n_t=0$. $A_{\rm lens}$ and $r$ are free parameters of the model that describe the residual lensing $B$-modes and the primordial $B$-mode fluctuations respectively. The $EE$ power spectrum, as well as $C_\ell^{{\rm lens},BB}$ and $C_\ell^{{\rm tens},BB}$, are fixed to those given by the best-fit cosmological parameters used by \cite{2018PhRvL.121v1301B}. Assuming no cosmic birefringence, the parity-violating $EB$ component of the CMB is set to zero. Note that this is a potentially strong assumption given the partial detection of a non-zero birefringence angle in the Planck data~\cite{minamibirefringe}, and the simultaneous measurement of birefringence and calibration of polarization angles (e.g. as done in \cite{2019arXiv190412440M}) should be studied in more detail.

$\left.C^{\nu\nu',pp'}_\ell\right|_{\rm FG}$ in Eq. \ref{eq:cl_sky_full} is the foreground contribution to the power spectrum. We parametrize this as:
\begin{equation}
\left.C^{\nu\nu',pp'}_\ell\right|_{\rm FG}=C^{S\times S,\nu\nu',pp'}_\ell+C^{D\times D,\nu\nu',pp'}_\ell+C^{S\times D,\nu\nu',pp'}_\ell,
\end{equation}
where $C^{c\times c,\nu\nu',pp'}_\ell$ is the auto-correlation of component $c$, and $C^{S\times D,\nu\nu',pp'}_\ell$ is the contribution from their cross-correlation.

The auto-correlations are modeled as
\begin{equation}\label{eq:cl_fg_auto}
C^{c\times c,\nu\nu',pp'}_\ell=f_c^\nu f_c^{\nu'}\,\left(\Delta_c\right)^{\log^2(\nu/\nu')/\log^2(\nu_{0,c})}\,C^{c,pp'}_\ell
\end{equation}
Here, $f^\nu_c$ is the frequency spectrum of $c$, which we model as a power law for synchrotron and a modified black body for dust \cite{2016A&A...594A...9P}. $\Delta_c$ is a decorrelation parameter, quantifying the decoherence of foregrounds as a function of frequency. Current constraints on $\Delta_D$ make it compatible with zero \citep{2018PhRvD..97d3522S}, and there is some evidence of non-zero $\Delta_S$ \cite{2018A&A...618A.166K}. Frequency decorrelation is one of the most relevant sources of systematic uncertainty for $B$-mode searches and can be caused, for example, by the spatial dependence of foreground spectral properties \citep{2017A&A...603A..62V}. The specific parametrization of frequency decorrelation used here corresponds to the model used by \cite{2018PhRvL.121v1301B}. The model is physically motivated, and corresponds to the form of decorrelation one would predict for Gaussian and scale-independent variations in the foreground spectral indices \cite{2017A&A...603A..62V}.  In antenna temperature units, the synchrotron and dust spectra are:
\begin{equation}
f^\nu_S=\left(\frac{\nu}{\nu_{0,S}}\right)^{\beta_s}, \hspace{12pt}
f^\nu_D=\left(\frac{\nu}{\nu_{0,D}}\right)^{\beta_d+1} \frac{e^{h\nu_{0,D}/k_B T_d} - 1}{e^{h\nu/k_B T_d} - 1},
\end{equation}
where $\beta_s$ and $\beta_d$ are synchrotron and dust spectral indices, and $T_d$ is the dust temperature. 

We parametrize the scale-dependent factor $C^{c,pp'}_\ell$ in Eq. \ref{eq:cl_fg_auto} as a power law of the form
\begin{equation}
\frac{\ell(\ell+1)}{2\pi}C^{c,pp'}_\ell\equiv A^{pp'}_c\left(\frac{\ell}{\ell_0}\right)^{\alpha^{pp'}_c},
\end{equation}
described by an amplitude $A^{pp'}_c$ and tilt $\alpha^{pp'}_c$. Note that we allow for non-zero foreground $EB$ components \citep{Keating2013,2016MNRAS.457.1796A,2019arXiv190412440M}. We choose a pivot scale $\ell_0=80$, and pivot frequencies $\nu_{0,S}=23\,{\rm GHz}$ and $\nu_{0,D}=353\,{\rm GHz}$.

The cross-correlations between components are modeled as:
\begin{equation}
C^{S\times D,\nu\nu',pp'}_\ell =\varepsilon_{DS}\,\left[f^\nu_Df^{\nu'}_S\sqrt{C^{D,pp}_\ell C^{S,p'p'}_\ell}+f^\nu_Sf^{\nu'}_D\sqrt{C^{S,pp}_\ell C^{D,p'p'}_\ell}\right].
\end{equation}
Here $\varepsilon_{DS}$ is the dust-synchrotron correlation coefficient. Note that we do not include decorrelation effects in the cross-correlation term \citep{2018PhRvL.121v1301B}. In total, our sky model is determined by the following set of 19 free parameters: 

\begin{multline*}
\{A_{\rm lens},\, r,\, A_S^{EE},\, A_S^{EB},\, A_S^{BB},\, \alpha_S^{EE},\, \alpha_S^{EB},\, \alpha_S^{BB},\, \Delta_S,\, \beta_S, \\
A_D^{EE},\, A_D^{EB},\, A_D^{BB},\, \alpha_D^{EE},\, \alpha_D^{EB},\, \alpha_D^{BB},\, \Delta_D,\, \beta_D,\, \varepsilon_{DS}\}
\end{multline*}

In general, we have assumed that the foreground fitting model is exactly the one used to generate the data. The parameters may differ in value, but the shape is correct (i.e.,  power law in, power law out). For data one would explore many foreground models, but for estimating the impact of systematics this method will capture the dominant systematic and foreground interactions.

\begin{table}
\centering
\begin{tabular}{| c | c  c  c  c  c  c  c |}
\hline
Parameter & $r$ & $A_{lens}$ & $A^{BB}_D$ & $\alpha^{BB}_D$ & $A^{EE}_D$ & $\alpha^{EE}_D$ & $\beta_D$ \\ 
Input Value & 0 & 1 & $20~\mu K^2$ & -0.2 & $40~\mu K^2$ & -0.4 & 1.53 \\
Prior & [-1, 1] & [0, 10] & $[0,\infty)$ & [-4, 1] & $[0,\infty)$ & [-4, 1] & $[0.1, 10]$ \\ \hline
Parameter & $A^{BB}_S$ & $\alpha^{BB}_S$ & $A^{EE}_S$ & $\alpha^{EE}_S$ & $\beta_S$ & $\varepsilon_{DS}$ & \\
Input Value & $5~\mu K^2$ & -0.6 & $10~\mu K^2$ & -0.8 & -3.1 & 0.2 & \\
Prior & $[0,\infty)$ & [-4, 1] & $[0,\infty)$ & [-4, 1] & $[-10, 0]$ & [-1, 1] & \\ \hline
Parameter & $\nu_{0,D}$ & $T_D$ & $\nu_{0,S}$ & & & & \\ 
Input Value & $353~GHz$ & $19.6~K$ & $23~GHz$ & & & & \\
Prior & const. & const. & const. & & & & \\ \hline
\end{tabular}
\caption{Summary of nominal sky signal parameters and modeling priors.} 
\label{tab:fgparams}
\end{table}

Our default foreground parameters, listed in Table~\ref{tab:fgparams}, were informed by the results of \planck \cite{planckxxx}, SPASS \cite{2018A&A...618A.166K}, and \bk \cite{2018PhRvL.121v1301B}, and selected on the brighter side of the ranges to be as conservative as possible. For dust we have a fixed temperature of $T_D=$19.6~K \citep{2015A&A...576A.107P}. The spectral index $\beta_D=1.53$. The power spectrum tilts are $\alpha^{EE}_D = -0.4$ and $\alpha^{BB}_D = -0.2$. The dust amplitudes are $A^{EE}_D=40 \mu K^2$ and $A^{BB}_D=20 \mu K^2$ at a reference frequency of $\nu_{0,D}=353$~GHz. For synchrotron, $A^{EE}_S=10 \mu K^2$ and $A^{BB}_S=5 \mu K^2$ at a reference frequency $\nu_{0,S}=23$~GHz. The spectral index is $\beta_S=-3.1$ and the power spectrum tilts are $\alpha^{EE}_S = -0.8$ and $\alpha^{BB}_S = -0.6$. The input synchrotron-dust correlation is $\varepsilon_{DS}=0.2$. We include no intrinsic foreground $EB$ component in the fiducial data vector. We use top-hat priors for all CMB and foreground parameters, chosen to be wide enough that the data completely constrains these parameters. In particular, foreground amplitudes must be positive, and foreground harmonic space tilts $\alpha_c$ are bound to [-4, 1]. $\varepsilon_{DS}$ is bound to [-1, 1]. The CMB priors are wide top-hats, $A_{lens}\in[0, 10]$ and $r\in[-1, 1]$. 

Our choice of foreground model corresponds to the minimal model required by current data, in addition to the possibility of foreground decorrelation. In terms of the requirements on systematics, we do not explore the wide variety of potential additional foreground models as this result should not be interpreted as a forecast. One of the primary drawbacks of our model is that lack of flexibility for spatially varying foregrounds. The decorrelation parameter can be interpreted as a model for Gaussian, spatially varying foreground spectral indices. Another promising method to account for spatial variations of the foregrounds while still performing an analysis in power-spectrum space would be to use the moment expansion introduced in~\cite{chlubahillabitbol2017} and applied in~\cite{mangilli2019,remazeilles2020}. The moment expansion relies on Taylor expanding the foreground SEDs into hierarchically ordered moments. While the decorrelation parameter can be related to the moments under the right assumptions, we do not consider the general case of additional foreground moments. The primary interaction of systematics will be on the first order foreground terms, all of which are included here. More complex foregrounds, or more foreground moments, will undoubtedly impact the $r$ constraint to some extent. As seen later in Section~\ref{ssec:results.std}, the interaction of decorrelation and systematics is subdominant to the systematics and decorrelation alone. In other words, additional terms arising from systematic cross moment parameter interactions should be even less significant than the systematic cross first-order foreground terms considered here. Lastly, as discussed in~\cite{chlubahillabitbol2017}, one could even interpret or combine systematic parameters with the foreground moments, which warrants further research. We therefore leave such considerations to future forecasting studies.

\subsection{Instrument model}
\begin{figure}[tbp]
\centering
\includegraphics[width=0.46\textwidth]{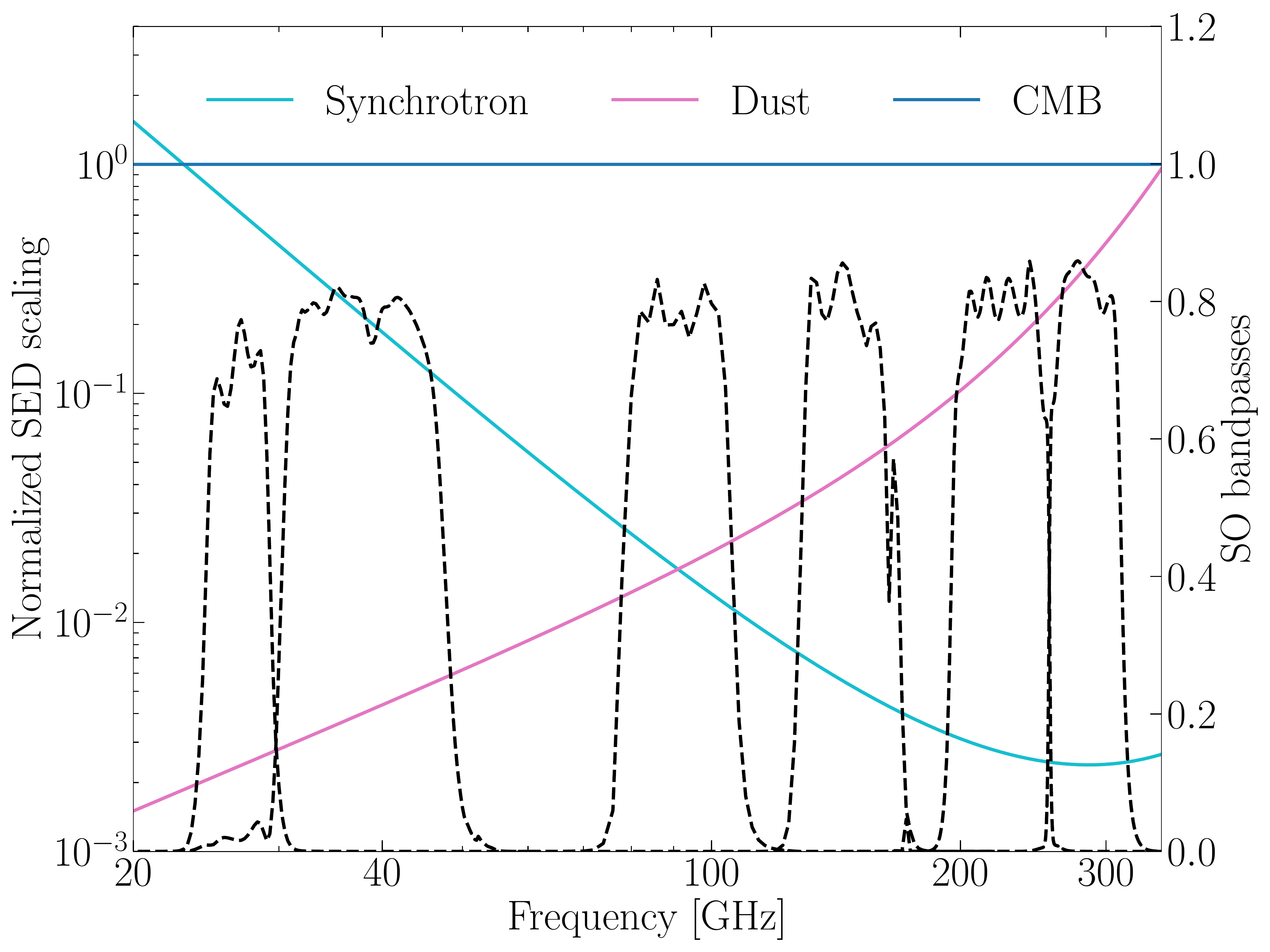} \hfill
\includegraphics[width=0.50\textwidth]{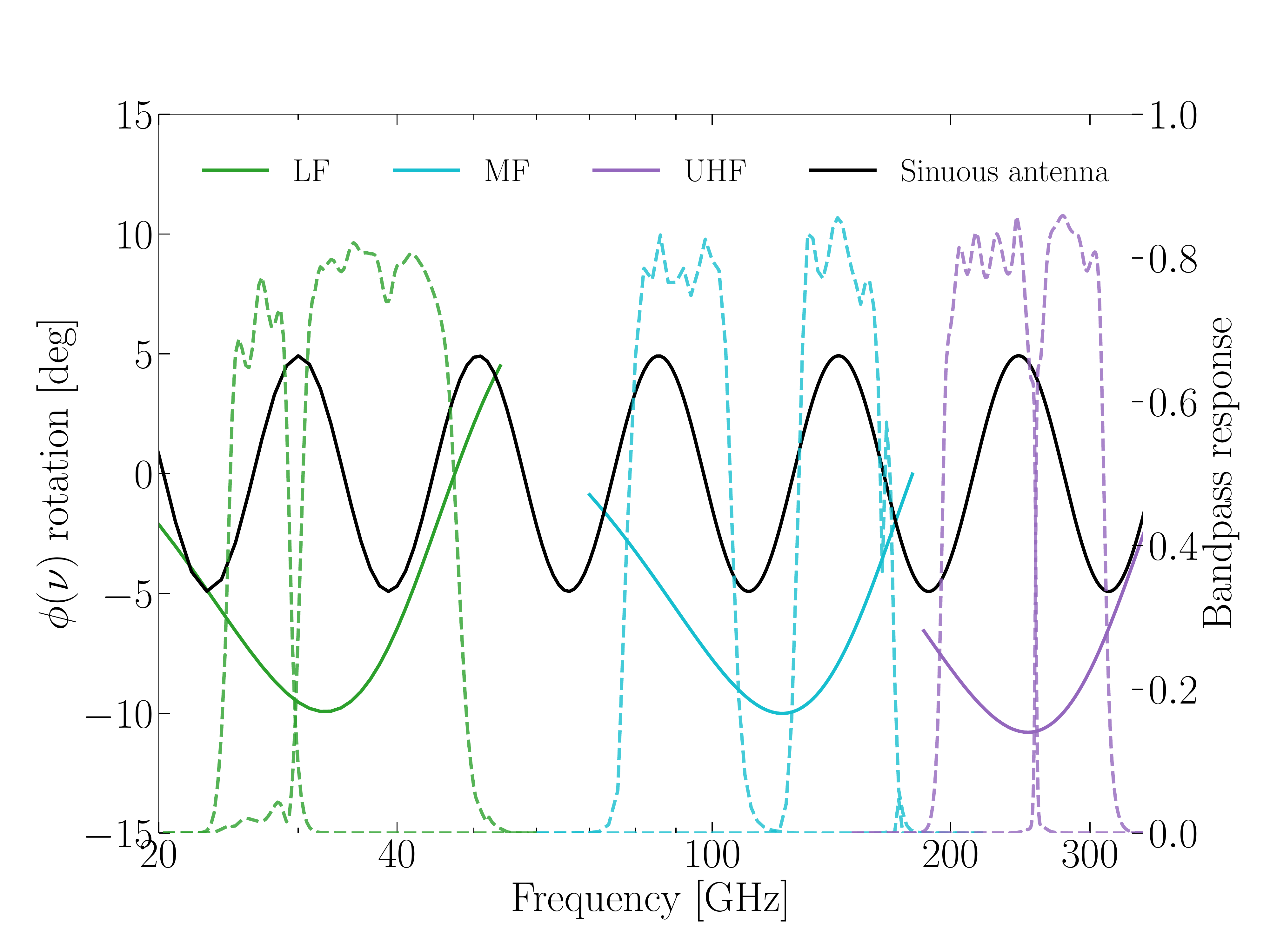}
\caption{(Left) SED scaling laws and simulated SO bandpasses. The signals are normalized to thermodynamic CMB units. The synchrotron scaling has a reference frequency (and value of unity) at $\nu_0=23$~GHz. The dust reference frequency is $\nu_0=353$~GHz. Note the SED and frequency axes are log scaled (while the bandpass axis is linear). (Right) Frequency dependent polarization angle rotations compared to SO bandpasses. The black curve represents the polarization rotation from a broadband sinuous antenna design (which is only planned for the LF bands in SO). The solid green, blue, and purple curves represent the frequency dependent rotation from a 3-layer HWP, for the LF, MF, and UHF bands. The simulated SO bandpasses are in dashed lines with the same corresponding colors. }
\label{fig:sedbandpass}
\end{figure}

The instrument systematics model is intended to describe uncertainties in the bandpasses and orientation of the polarization angle. All systematics modeled here are assumed to be constant across the map. We do not model variations across time or focal plane position. Instead, our model aims to parametrize the effective residual systematics remaining in the final coadded maps. This study could be extended in the future to explore the temporal and spatial variation of these systematics making use of time-ordered-data simulations, or the analytical methods recently proposed by \cite{2020arXiv200800011M}. We assume the bandpasses (and all systematics) are the same in $Q$ and $U$ (or $E$ and $B$) in the map domain. 

Similar studies for satellite missions such as LiteBIRD have explored the requirements for both per-detector and band-averaged bandpass systematics~\cite{ghigna2020}. In particular, \cite{ghigna2020} used map-domain simulations to find that the per-detector calibration requirements are approximately an order of magnitude less restrictive than the band-averaged results. Additionally, they study the impact of finite resolution bandpass measurements. Broadly, our results are in agreement with their findings. For example, as we will present below, both studies agree that the higher frequency bands have the most strict calibration requirements, at approximately the sub-percent level, although a direct comparison is not immediately possible due to the different experimental configurations. 

\subsubsection{The SO Small Aperture Telescopes}
The SO Small Aperture Telescopes are a set of three 42~cm-aperture telescopes containing a total of $\sim30,000$ detectors operating at 100 mK. Each telescope contains seven detector arrays, with two SATs operating at 94 GHz and 148 GHz (mid-frequencies, MF), and one SAT observing at 225 and 280 GHz (ultra-high frequencies, UHF) for a nominal period of five years. An additional optics tube observing in two low-frequency  bands (LF, 27 and 40 GHz) will be deployed for a single year and then replaced by an MF tube. Each telescope uses an achromatic sapphire continuously-rotating half-wave plate for polarization modulation. By mitigating intensity-to-polarization leakage, this minimizes the impact of the long time-scale fluctuations in the unpolarized atmosphere ($1/f$ noise), which increases polarization sensitivity at large angular scales where primordial B-modes are expected to peak~\cite{kusakaModulationCMBPolarization2014}. SO plans to use sinuous antennas with lenslets for the LF detector optical coupling, and orthomode transducers (OMT) with feedhorns for the MF and UHF bands~\cite{wobble,OMT_2009}.

To model the signal in each frequency band, we use simulated bandpass transmission curves. These include the simulated bands including dielectric losses from the preliminary on-chip stub filter designs and circuitry calculated with Sonnet\footnote{https://www.sonnetsoftware.com/} and the transmission through the optical coupling structure modeled in High Frequency Structure Simulator (HFSS)\footnote{https://www.ansys.com/products/electronics/ansys-hfss}. We note that the SO LF bands were not finalized at the time of this work, so we use the simulated LF Advanced ACTPol (AdvACT) bands for this analysis~\cite{Simon_2018}. These bandpasses are meant to provide representative functions of the array averaged bandpasses. These are shown in Figure~\ref{fig:sedbandpass}.

We model the angular resolution of each band through a Gaussian beam with a diffraction-limited full-width at half maximum, corresponding to $\theta_{\rm FWHM}=$ (91, 63, 30, 17, 11, 9) arcmin for the 6 different frequency bands (from low to high frequencies). The beam-corrected noise spectra, together with the expected signal are shown in Figure \ref{fig:noiseps}. We assume an observed sky fraction $f_{\rm sky}=0.1$.

\subsubsection{Bandpass systematics}
We model uncertainties in the bandpasses through two parameters, describing an effective shift in the mean frequency, $\Delta\nu$, and a change in the end-to-end calibration factor, which we will call a gain or calibration difference $\Delta g$. The total on-sky bandpass includes both the transmission through the telescope and through the atmosphere. Uncertainties in the bandpass can be sourced by a number of effects. These include fabrication variation in the on-chip bandpass filters between both detectors and wafers, time variation from atmospheric fluctuations, and systematic effects in their measurements with a Fourier transform spectrometer (FTS). This simple parametrization encapsulates most typical bandpass uncertainties without adding extra parameters. The bandpass shift is a commonly used parameterization to absorb bandpass uncertainties~\cite{2018ApJ...861...82W}. Nevertheless, it is worth emphasizing that, at the end of the day, what is needed is a full characterization of the bandpass shape, its uncertainties, and how those propagate into uncertainties in the corresponding central frequencies.

For example, increasing the width of a band would be expressed as a change in the center frequency and gain of the band. We do not explicitly consider the interaction of bandpass shifts with the atmospheric transmission lines, but the variation in the bandpass with the atmosphere can be parametrized by a center frequency shift and gain variation, so this study could give a limit on the allowable level of atmospheric variation in future work.

We assume that a given frequency band $b$ has been characterized by a measured bandpass transmission curve\footnote{Note that we use transmission curves defined in terms of intensity (e.g. as in \cite{2014A&A...571A...9P}) as opposed to antenna temperature.} $W_b(\nu)$. Assuming no systematics, ignoring for now the effects of a finite telescope beam, and assuming bandpass transmission curves are measured with respect to a Rayleigh-Jeans source, a map of the sky in CMB temperature units in that band would therefore be given by
\begin{equation}\label{eq:bandpass_int}
{\bf m}_b(\vec{\theta}) = N_b^{-1}\,\int d\nu\,\nu^2\,W_b(\nu) {\bf m}_\nu(\vec{\theta}),\hspace{12pt}
N_b\equiv\int d\nu\nu^2 W_b(\nu)\,f_C^\nu,
\end{equation}
where ${\bf m}_\nu(\vec{\theta})\equiv(Q_\nu(\vec{\theta}),U_\nu(\vec{\theta}))$ is the sky emission as a function of frequency. In the presence of bandpass systematics, the observed map becomes
\begin{equation}\label{eq:map_bands}
{\bf m}_b(\vec{\theta}) = N_b^{-1}\,(1+\Delta g_b)\int d\nu\,\nu^2\,W_b(\nu+\Delta\nu_b) {\bf m}_\nu(\vec{\theta}),
\end{equation}
where $\Delta g_b$ is the error in gain calibration and $\Delta nu_b$ is a frequency shift, for a given band $b$. The effect on the cross-frequency power spectrum is then
\begin{equation}\label{eq:cl_bands}
C^{bb',pp'}_\ell= \int d\nu\,\nu^2\,\frac{1+\Delta g_b}{N_b}W_b(\nu+\Delta\nu_b)\int d\nu'\,\nu'^2\,\frac{1+\Delta g_{b'}}{N_{b'}}W_{b'}(\nu'+\Delta\nu_{b'})\, C^{\nu\nu',pp'}_\ell .
\end{equation}

\subsubsection{Polarization angle systematics}
In addition to uncertainties on bandpass properties, we also study the impact of polarization angle uncertainties, which we parametrize as a frequency-dependent phase $\phi_b(\nu)$. In the presence of such a frequency-dependent angle, Eq. \ref{eq:map_bands} is modified by multiplying ${\bf m}_\nu(\bar{\theta})$ by the rotation matrix:
\begin{equation}
{\sf R}(\phi_b(\nu))\equiv
\left(
\begin{array}{cc}
\cos2\phi_b(\nu) & -\sin2\phi_b(\nu) \\
\sin2\phi_b(\nu) & \cos2\phi_b(\nu)
\end{array}
\right)
\end{equation}
The effect on the power spectra in Eq. \ref{eq:cl_bands}
is:
\begin{equation}\label{eq:cl_bands_pol}
C^{bb',pp'}_\ell= \sum_q\int d\nu\,\nu^2\,{\cal W}^{pq}_b(\nu)\sum_{q'}\int d\nu'\,\nu'^2\,{\cal W}^{p'q'}_{b'}(\nu')\,C^{\nu\nu',qq'}_\ell,
\end{equation}
where $p$ and $q$ are indices of the polarization angle rotation matrix and we have defined:
\begin{equation}\label{eq:bandpass_angle}
{\cal W}^{pq}_b(\nu)\equiv \left[{\sf R}(\phi_b(\nu))\right]^{pq}\,\frac{1+\Delta g_b}{N_b}W_b(\nu+\Delta \nu_b).
\end{equation}

We will consider the following cases for $\phi_b(\nu)$:
\begin{itemize}
\item {\bf Achromatic half-wave plate (HWP):} As described in further detail in Section~\ref{sssec:css.design.hwp}, the half-wave plate's sapphire layers cause a frequency-dependent polarization angle rotation \citep{2012ApJ...747...97B}. This can be mitigated by stacking several layers of sapphire, minimizing polarization angle variation and improving polarization modulation efficiency at the expense of mechanical risk and additional costs. We therefore explored the systematic effects associated with the frequency-dependent polarization angle of a minimal 3-layer HWP. The corresponding frequency-dependent angle for the LF, MF and UHF bands in the SO SAT, calculated using a Mueller matrix approach (see e.g. \citep{2020arXiv200907814V}), are shown in Figure~\ref{fig:sedbandpass} in solid blue, orange and green respectively.
\item {\bf Sinuous antennas:} The sinuous antennas have a frequency-dependent polarization angle wobble. Figure~\ref{fig:sedbandpass} shows the polarization angle variation modeled in HFSS for a design similar to the POLARBEAR2 design across the full range of frequencies~\cite{wobble} (see \cite{suzuki_thesis} for specific details of the model). We note that the frequency of the wobble can be increased and the amplitude of the wobble can be decreased by increasing the density of the switchbacks in the sinuous antenna design~\cite{wobble}, which is currently under study for the SO LF detectors. Additionally, the impact of the wobble could be greatly mitigated by populating the focal plane with sinuous antennas of opposing handedness. Sinuous antennas with opposite handedness will have polarization rotations with the same amplitude but in the opposite direction. With proper calibration of the detector bandpasses, the total polarization wobble effect could potentially be reduced by combining maps from two opposite handed sinuous antennas~\cite{suzuki_thesis}. Systematic mis-calibration of bandpass and detector gain between detectors with opposite handedness will result in a residual polarization rotation effect, which could then be parametrized by the generic parametrization method described below.

\item {\bf Generic parametrization:} Lastly, we attempt to capture the main effects of a slow variation of the polarization angle across the band through a two-parameter, first-order Taylor expansion:
\begin{equation}\label{eq:phinu_taylor}
\phi_b(\nu)=\Delta\phi_{0,b} + \Delta\phi_{1,b}\frac{\nu-\bar{\nu}_b}{\bar{\nu}_b},
\end{equation}
where $\bar{\nu}_b$ is the mean frequency of the band for the CMB spectrum. This linear parametrization of the frequency dependence can be used to capture unmodeled effects from the HWP or sinuous antennas.
\end{itemize}

Note that the impact of a frequency-dependent polarization angle in $B$-mode searches is two-fold in the presence of foregrounds: first, if uncorrected, a residual polarization angle miscalibration causes $E$-$B$ leakage, therefore giving rise to a spurious $B$-mode signal and a biased estimate of $r$. Secondly, if the frequency dependence is not trivial, the effect on the polarized emission of different components (CMB, dust and synchrotron), will be different due to their different spectra. If the effect is sufficiently large, this would require separate calibrations of the polarization angle for the different spectra or, more optimally, for the effects of the frequency-dependent angle to be forward-modeled within the component separation stage. Additionally, it is worth noting that polarization angle systematics can be correlated with bandpass uncertainties, particularly if the frequency dependence of the polarization angle is steep within the band.

In the case of a $\phi_b(\nu)$ associated with HWPs and sinuous antennas, we will explore two scenarios. First, we will consider the case where the frequency dependence of the angle is well-known and can be corrected for at the map level. In this case, we include the rotation angle as part of the bandpass ${\cal W}_b$ as in Eq. \ref{eq:bandpass_angle} when computing the bandpass integral normalization ($N_b$ in Eq. \ref{eq:bandpass_int}). In this case, the effect of the frequency-dependent angle is completely corrected assuming a CMB source spectrum. A residual effect will still remain due to the different spectra of the other components. This can either be ignored if it is small enough compared to the statistical uncertainties on $r$, or taken into account in component separation. We will consider the former case, and therefore quantify whether the spectral response of $\phi_b(\nu)$ to different sources leads to a significant bias on $r$.

Secondly, we will assume that the frequency dependence of $\phi_b(\nu)$ is completely unknown, and we will try to account for it within the analysis pipeline using the first-order expansion in Eq. \ref{eq:phinu_taylor}. This allows us to explore whether this approximate model is able to accurately reproduce realistic frequency-dependent angle variations.

Finally, we will use the first-order expansion to place requirements on the calibration of the overall polarization angle in each band ($\Delta\phi_{0,b}$) and on its frequency slope ($\Delta\phi_{1,b}$).

We explored a variety of priors on systematics. Most of our results use $\pm 3\%$ top-hat priors for the gain and shift parameters, unless otherwise noted. $\Delta\phi_1$ is bound to [$-45^{\circ}$, $+45^{\circ}$], corresponding to an edge-to-edge variation of the polarization angle of $\approx 15^{\circ}$. In general, the systematic priors are wider than what is usually achieved with fielded instruments, and therefore should serve as a conservative range.

\section{Results}
\label{sec:results}

\subsection{Requirements on unmarginalized systematics}\label{ssec:results.bias}
We start by exploring the level to which the bandpass and polarization angle systematics described in the previous section must be known in order to avoid a significant bias on the final constraints on $r$.

To do so, we must first determine what an acceptable bias on $r$ would be. In the case of SO, the expected 1$\sigma$ error on $r$ is $\sigma(r)\simeq2\times10^{-3}$, with ${\cal O}(1)$ variations depending on the choice of component separation method and noise model. Since we will explore biases from the effects of a large number (${\cal O}(10)$) of systematic parameters, we will impose the conservative criterion that no individual parameter should induce a bias larger than $1/10$ of the expected statistical uncertainties (i.e. $\Delta r=2\times10^{-4}$). As a less conservative threshold, and for purposes of comparison, we will also report the calibration requirements associated with a bias $\Delta r=10^{-3}$.

\begin{figure}[tbp]
\centering
\includegraphics[width=0.48\textwidth]{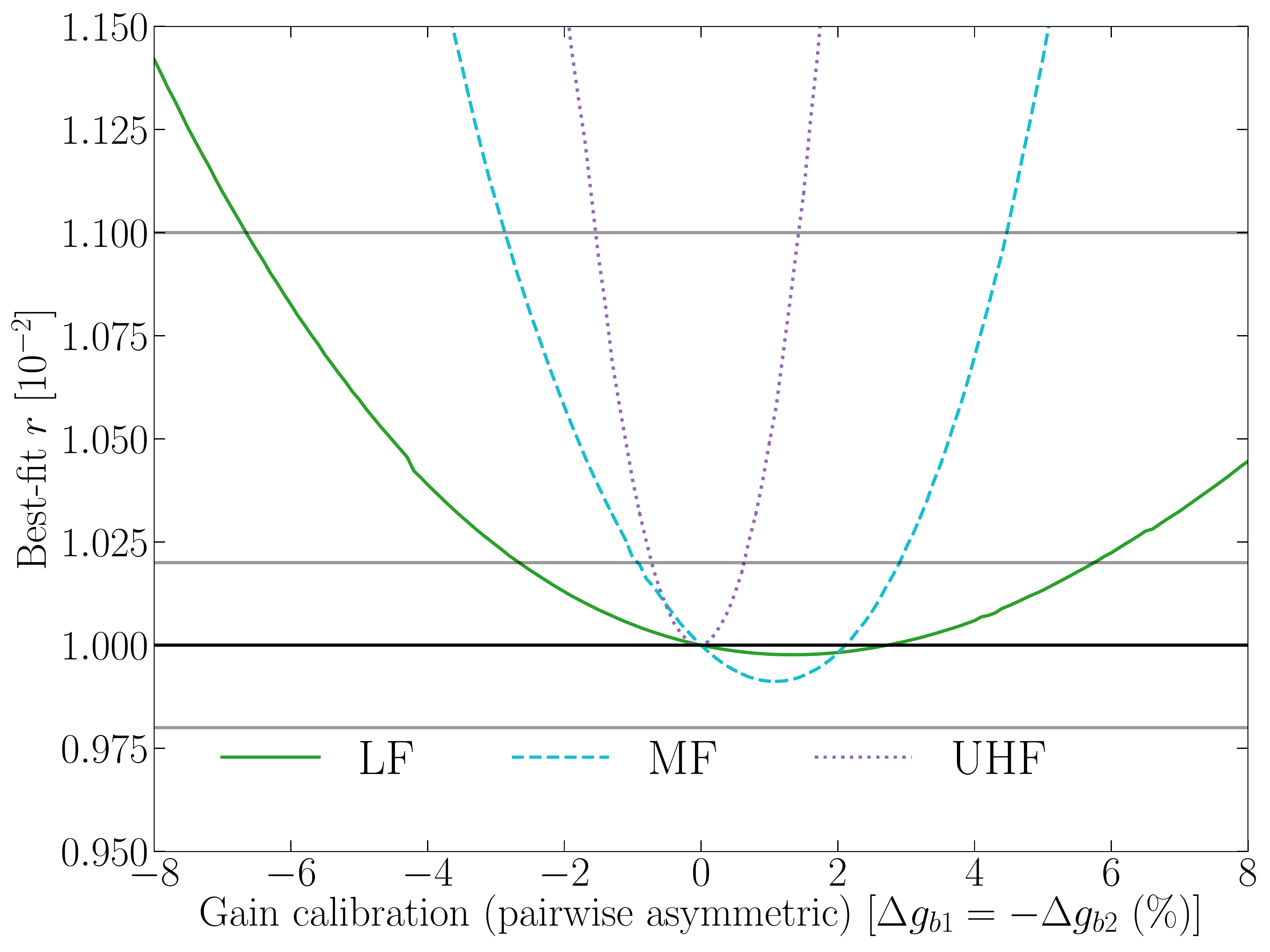} \hfill
\includegraphics[width=0.48\textwidth]{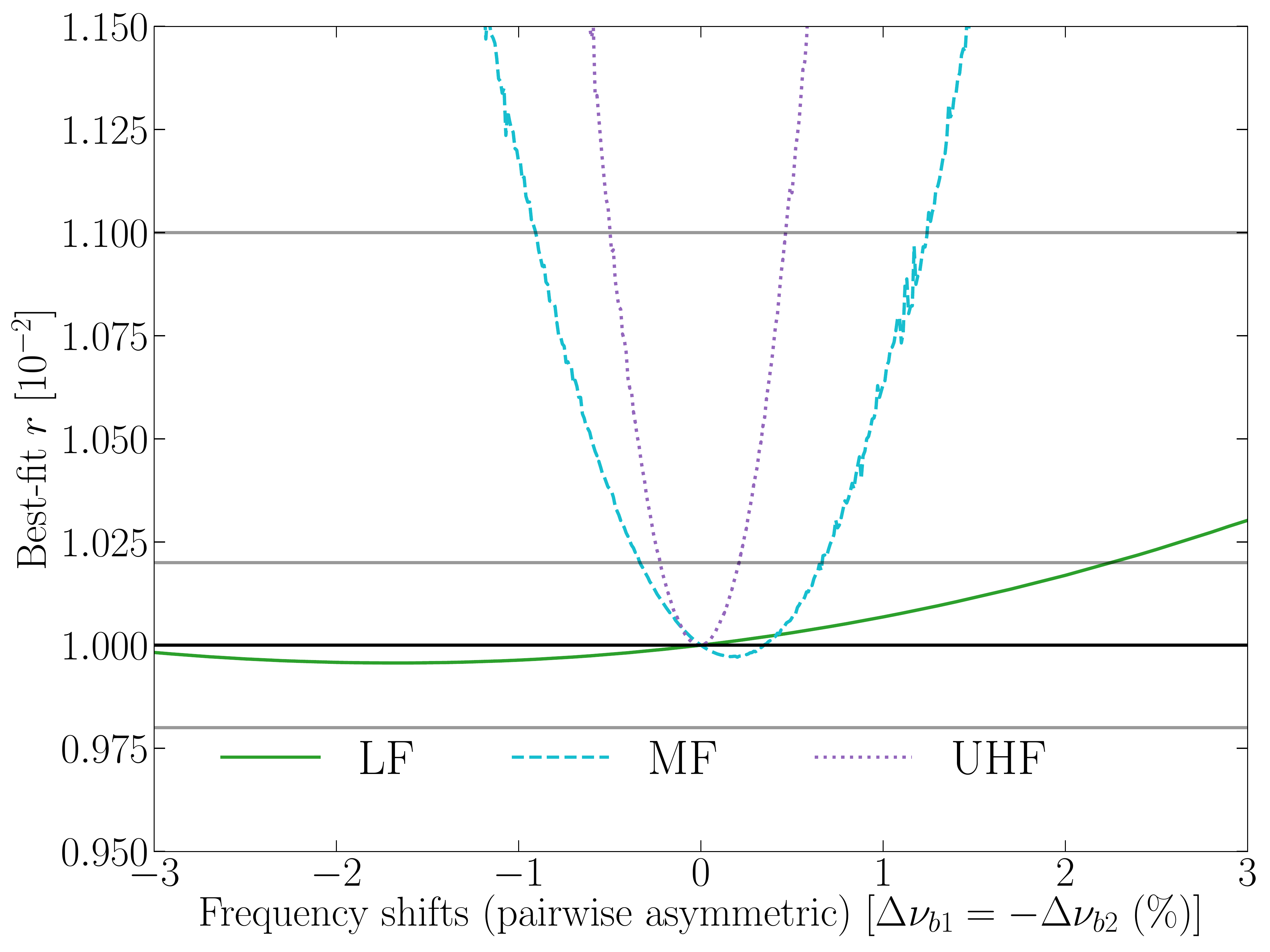} \\ 

\includegraphics[width=0.48\textwidth]{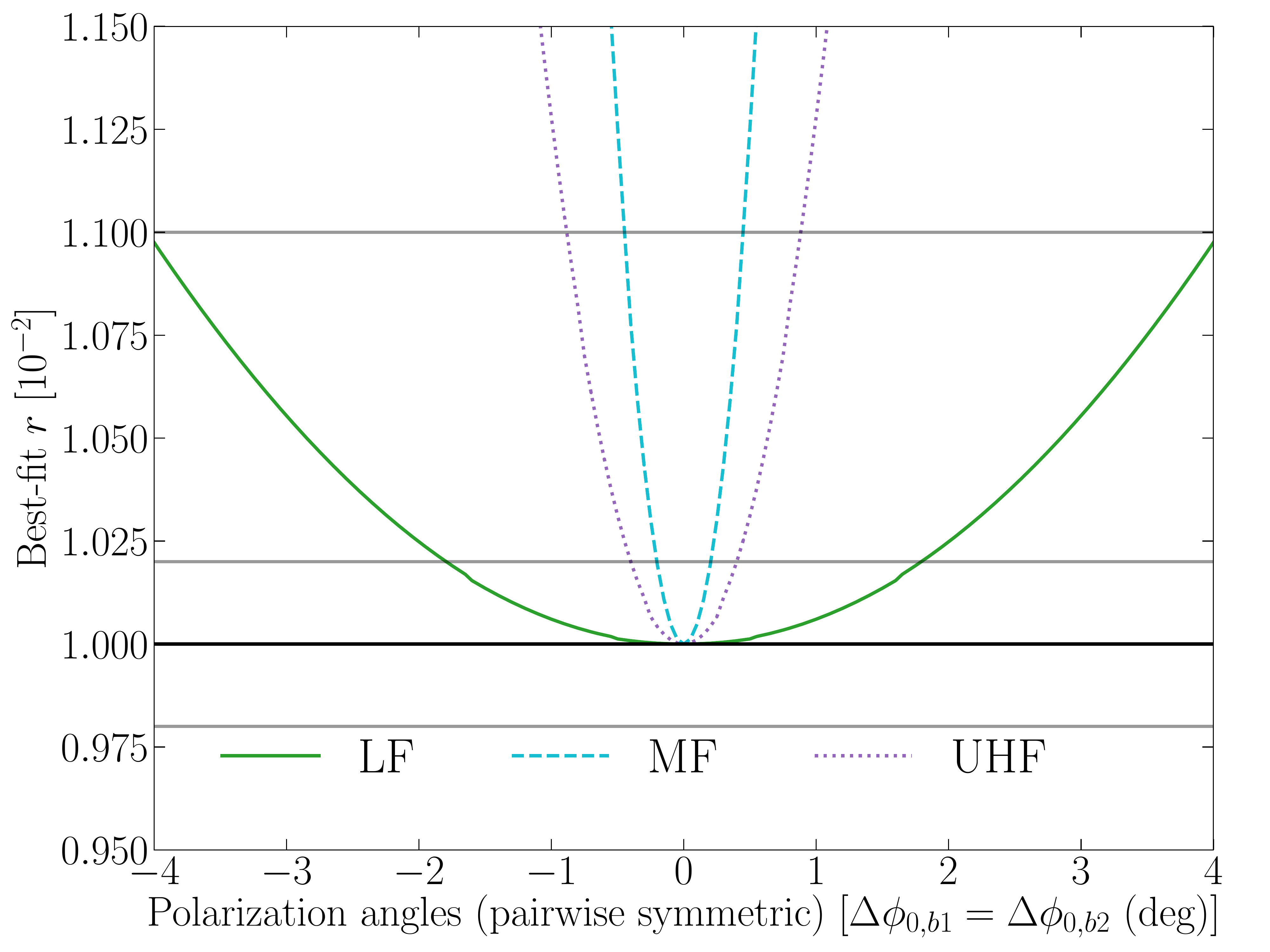} \hfill
\includegraphics[width=0.48\textwidth]{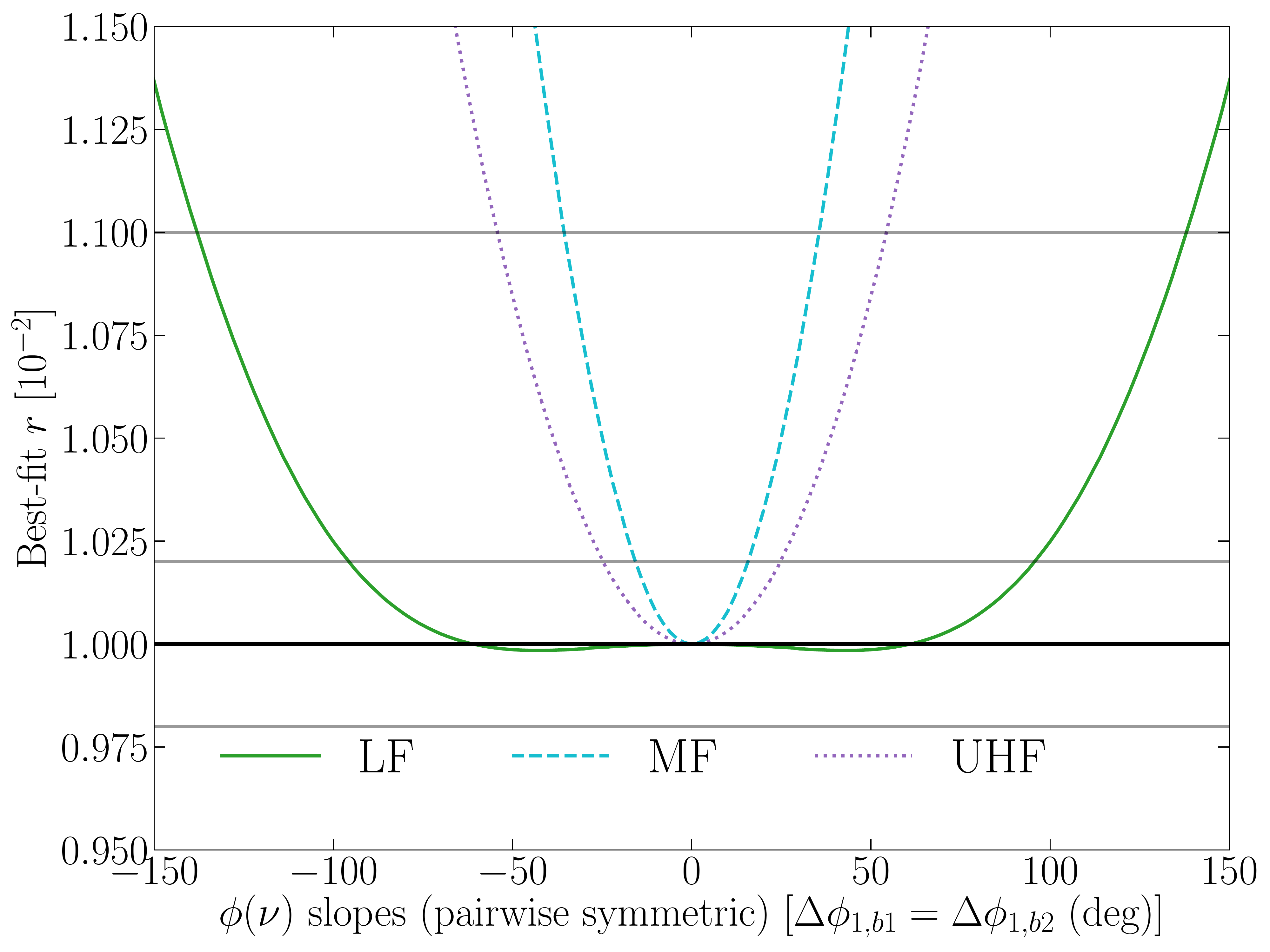}
\caption{{\bf Top:} Bias on $r$ as a function of the difference in gain calibration (left) and frequency shifts (right). Both top panels show antisymmetric variations (i.e., the parameter values change by the same amount but in opposite directions for the two frequency bands in a pair) for the LF (solid, green), MF (dashed, blue) and UHF (dotted, purple) frequency pairs. In both cases, the tightest requirement, corresponding to $\sim$percent-level calibration of both systematics, is found for the UHF band. The requirements on the LF band, on the other hand, are a factor $\sim$3 laxer, owing to the reduced sensitivity and angular resolution of that band. {\bf Bottom:} Bias on $r$ as a function of the polarization angle systematic parameters $\Delta\phi_0$ (left) and $\Delta\phi_1$ (right), corresponding to the offset and frequency slope of a linear frequency-dependent polarization angle (see Eq. \ref{eq:phinu_taylor}). Both bottom panels show symmetric variations of the parameters (i.e., the parameter values change by the same amount in both frequency bands in a pair) for the LF, MF and UHF band pairs. In both cases, the tightest requirement is found for the MF band. This is due to the large amplitude of CMB $E$-modes leaked into $B$-modes, which is more relevant for the MF band, while the LF/UHF bands have a lower weight in the final cosmological constraints.}. The requirements on the LF band are significantly more lax, due to the reduced sensitivity and angular resolution of that band. In all the plots we also show the input $r=0.01$ and thresholds ($\Delta r=10^{-3}$ and $\Delta r=\pm2\times10^{-4}$) used to quantify the requirements on both parameters, as thin horizontal black lines.
\label{fig:reqs_bpass}
\end{figure}

The calibration requirements on the systematic parameters are calculated as follows. We fix the input data to the nominal sky model described in Section \ref{ssec:smodel}, which include no systematics (i.e.,  all systematic parameters are zero). We then fit our sky model to these data, setting some of the systematic parameters in the model to non-zero values, while fixing all other systematic parameters and varying only the CMB and foreground parameters. In reality, the opposite occurs, and the systematics should go on the data side not the model side of course. Choosing to apply the systematic on the model instead of the data allows us to easily execute many minimization runs in parallel and explore the impact of a given systematic in detail. Although the minimization equation is not exactly symmetric in this way, due to small changes in the covariance matrix, we confirmed that this is a good approximation by comparing a handful of cases to the more realistic scenario where the systematics are applied to the data. The results agreed to better than 2\%.

Our setup includes four systematic parameters in each of the six frequency bands for a total of 24 dimensions, which would make a true grid search impractical (just sampling each systematic twice would incur $2^{24}\simeq1.6\times10^7$ minimizations). Instead, we approach this by exploring each systematic (bandpass shifts, gains, average polarization angles, and frequency-dependent angles) one at a time. We vary each systematic on a band-by-band basis (i.e.,  varying its value in one single frequency band at a time). Since the LF, MF, and UHF pairs of bands will be on sky at the same time, we also explore pairs of systematics which we call {\sl symmetric} and {\sl antisymmetric} variations in the systematic parameter values of the two bands in a pair. For {\sl symmetric} variations, both parameters take the same value, while in {\sl antisymmetric} variations they take opposite values away from the center value of zero. Although we did not explore the full parameter space, this should be a fairly conservative scenario, as we have intentionally created significant residuals with these systematic combinations. Note that coherent systematics across all bands will likely lead to more stringent constraints.

The worst-case scenario (i.e.,  the mode of variation yielding the maximum bias on $r$) is different for each systematic. Mismodeled systematics in a single frequency band always incur smaller biases on $r$ than mismodeled systematic parameters in paired frequency bands, so we report the constraints from the paired systematic scenarios. We find that antisymmetric variations lead to the most stringent requirements on gain calibration uncertainties and bandpass frequency shifts. This is because symmetric changes in the gain calibration or frequency center, primarily impact the amplitudes of the foregrounds, while asymmetric variations impact the foreground SED spectral index. Incorrect SED spectral indices generate larger biases on $r$ than incorrect foreground amplitudes because the extrapolation across frequencies relies on the spectral index in a non-linear way.

Polarization angle systematics ($\Delta\phi_{0,1}$) are marginally more sensitive to symmetric variations. The rotation equations for auto-correlations are symmetric in changes of parity of the angles, except for the $EB$ spectra and therefore the two scenarios lead to similar results.

The results for these worst-case scenarios are shown in the top panels of Figure~\ref{fig:reqs_bpass} for frequency shifts and gains, and in the bottom panels of Figure \ref{fig:reqs_bpass} for polarization angle systematics. The corresponding requirements are listed in Table~\ref{tab:syst_req_table}. 

We also compute a {\sl synthetic} probability-to-exceed (PTE) for these biases. We call them synthetic because the data are noise-less and so best-fit $\chi^2=0$, which would normally correspond to a PTE of 1. Instead, the synthetic PTE is calculated using $\chi^2_{\rm eff}=\chi^2+d.o.f$ so that a perfect model would have a PTE=0.5, and all deviations away from that will have PTE<0.5. This synthetic PTE is used to determine whether a systematic would go unnoticed, or whether the fit would find a poor $\chi^2$ and thus indicate a poorly fitting model. PTEs close to 0.5 show an undetected systematic combination that is degenerate with foregrounds in such a way that it might go unnoticed, resulting in a biased claim on parameters. PTEs near zero are highly detected and therefore indicate a poor fit, which would warrant further inspection.

The tightest requirements on bandpass systematics are found for the UHF band, corresponding to $\lesssim 1\%$-level calibration of both mean frequencies and gains. This requirement could pose a challenge for next-generation experiments \cite{2018ApJ...861...82W}. However, two mitigating factors must be noted: firstly, as shown in detail in Section \ref{ssec:results.std}, and hinted at in \cite{2018ApJ...861...82W}, a larger level of uncertainty on these parameters can be marginalized over without damaging the constraints on $r$ (see \citep{Bao:2015eaa}). Secondly, we quote the most-conservative requirements, found for asymmetric variations. However, since bandpasses are measured and calibrated using the same FTS, symmetric variations, for which the corresponding requirements are a factor $2-3$ larger, would arguably be a better motivated model. Additionally, as noted by the PTE values in the table, the residuals generated by these systematics are large enough to be detected with a poor $\chi^2$. The only undetected bandpass systematics are LF $\Delta g$ with a synthetic PTE of 0.17, and LF, UHF $\Delta\nu$ to a lesser extent.  

\begin{table}
\centering
\begin{tabular}{| c | c | c | c | c | }
\hline
Band & $\Delta g_b$ ($\%$) & $\Delta\nu_b$ ($\%$) & $\Delta\phi_{0,b}$ (deg) & $\Delta\phi_{1,b}$ (deg) \\ \hline
 \multicolumn{5}{|c|}{$\Delta r = 10^{-3}$} \\
\hline
LF & 6.6 [0.00] & 6.70 [0.00] & 4.05 [0.00] & 136 [0.00] \\
MF & 2.8 [0.00] & 0.90 [0.00] & 0.40 [0.00] & 34 [0.01] \\
UHF & 1.4 [0.00] & 0.46 [0.00] & 0.85 [0.00] & 54 [0.00] \\
\hline
\hline
\multicolumn{5}{|c|}{$\Delta r = 2\times10^{-4}$} \\
\hline
LF & 2.6 [0.17] & 2.20 [0.08] & 1.75 [0.10] & 94 [0.00] \\
MF & 0.9 [0.00] & 0.33 [0.0] & 0.20 [0.01] & 14 [0.40] \\
UHF & 0.6 [0.01] & 0.20 [0.06] & 0.40 [0.01] & 24 [0.07] \\
\hline
\end{tabular}
\caption{Systematic uncertainty requirements for a given maximum bias on $r$, with synthetic probability-to-exceed (PTE) in brackets. To measure the bias on $r$, data is generated using only the CMB and foregrounds, but the model includes the given systematic effect. The bounds are calculated using pairs of bands (e.g.,  the MF band refers to the 94 and 148~GHz bands). For the gain and shift parameters, the most conservative bounds come from antisymmetric combinations of the paired-bands, while the angle parameters use symmetric combinations of the systematic. For example, we find the UHF shift parameters need to be known better than 0.2\% for a bias of less than $2\times10^{-4}$, where one band was frequency shifted by -0.2\% and the other by +0.2\%. In the case of the angles, both bands are assigned the same value of the systematic. The PTEs indicate that most of these biases would be significantly detected by studying the residuals. The most problematic systematic biases are LF $\Delta g$ (PTE=0.17), LF $\Delta\phi_0$ (0.10), and MF $\Delta\phi_1$ (0.40).}
\label{tab:syst_req_table}
\end{table}

For polarization angle systematics, the tightest requirements are found for MF band, due to the large amplitude of CMB $E$-modes that leak into $B$-modes. The overall angle offset $\Delta\phi_0$ must be calibrated at the sub-degree level. This is a stringent requirement, but one that is achievable with current technology \cite{TauA}\footnote{Note however that future experiments with more stringent requirements will likely reach the current model error on the Tau-A polarization direction in terms of absolute calibration.}. The requirement on the frequency slope of the polarization angle $\Delta\phi_1$ is significantly softer. If phrased in terms of the difference in polarization angle between the two edges of the frequency band, the required calibration level is of the order of $\sim 5^\circ$ in the worst-case scenario. This result can be interpreted as follows: for a perfectly flat spectrum and bandpass, a linear tilt in the frequency dependence of the polarization angle has a net-zero effect. Therefore $\Delta\phi_1$ can only cause a bias through the frequency variation of the different SEDs across the band, which is relatively mild for all the sky components under consideration. Again the PTEs for the angle systematics indicate that residuals would be statistically significant and therefore would motivate further investigation. The MF $\Delta\phi_1$ poses the biggest problem here as the bias goes undetected with a PTE=0.4. 

\begin{figure}[tbp]
\centering
\includegraphics[width=0.6\textwidth]{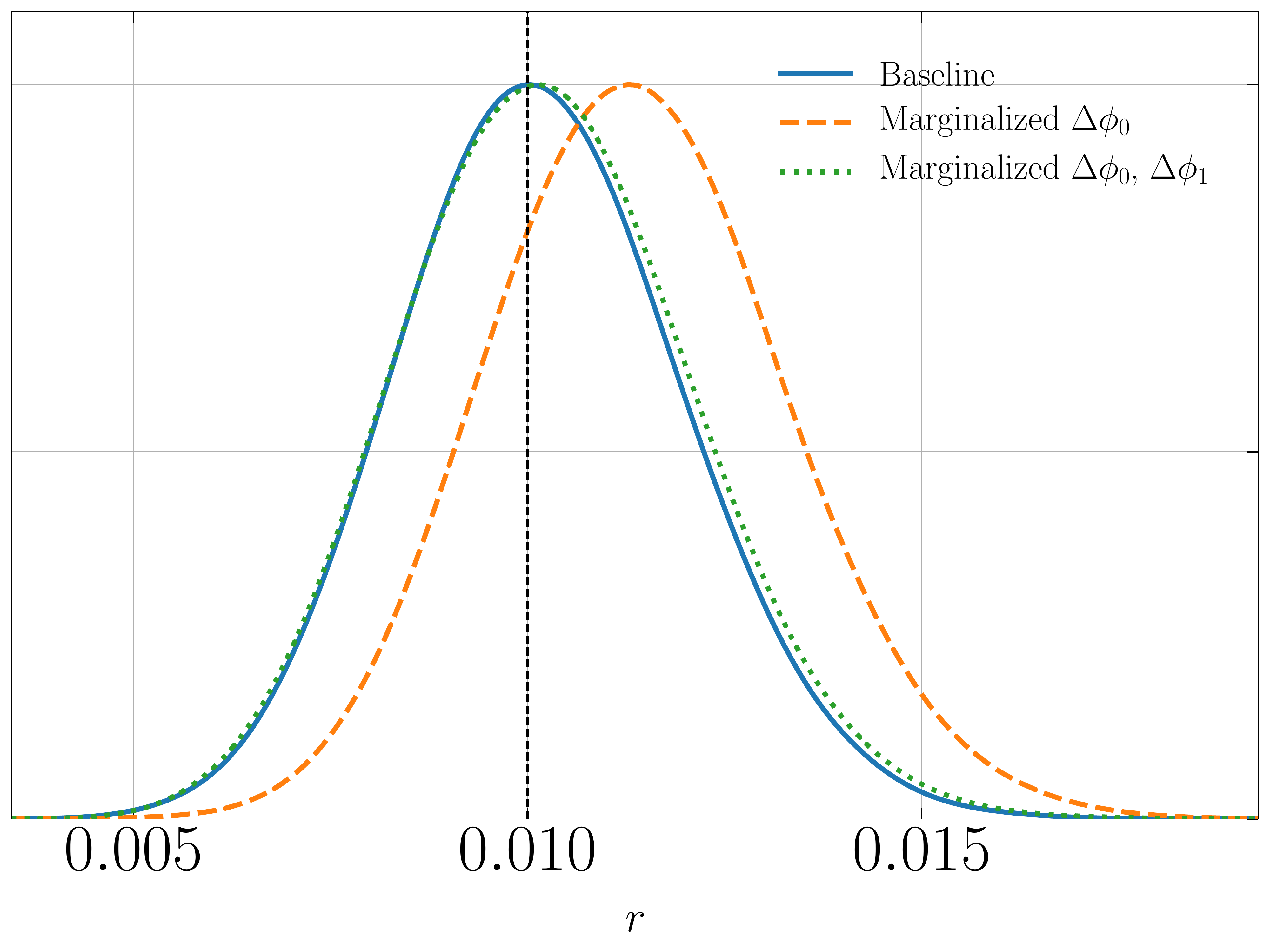} \hfill
\caption{Posterior on $r$ for different modeling assumptions of the frequency dependent rotation due to sinuous antennas. The baseline $\sigma_r=1.8\times10^{-3}$. Fitting for the CMB and FG model but ignoring $\phi(\nu)$ leads to $\Delta_r = 1.2\times 10^{-3}$ bias on $r$ (orange). If $\phi(\nu)$ is known, one can include in the model and completely eliminate bias (blue). If $\phi(\nu)$ is not known or there is residual $\phi(\nu)$, we can model and marginalize over it. Here we used a linear approximation with parameters $\Delta\phi_0$ and $\Delta\phi_1$ over each bandpass as a model for the sinuous $\phi(\nu)$. This eliminates the bias on $r$ ($\Delta_r=7\times10^{-5}$), with a negligible increase in $\sigma_r=1.9\times10^{-3}$ (green). }
\label{fig:sinuous}
\end{figure}

We have also studied the impact of the frequency-dependent polarization angles caused by the HWP and sinuous antennas. If ignored, the sinuous antenna angle results in a bias on $r$ of $\Delta r = 1.2\times10^{-3}$, even when allowing a constant angle parameter $\Delta\phi_0$ to vary. The 3-layer AWHP induced rotation produces a bias of $\Delta r = 1.5\times10^{-4}$. However, both of these biases can be almost completely eliminated by marginalizing over the additional $\Delta\phi_1$ parameter, which removes the leading order component of the frequency-dependent rotation. Additionally, this marginalization only increases $\sigma_r$ by $<10^{-4}$, as seen in Figure~\ref{fig:sinuous}. 

If the frequency dependence of the polarization angle is well known, we should be able to correct for it exactly, assuming perfect knowledge of the spectra of all sky components. To explore the sensitivity of this effect to uncertainties in the component separation model, we have quantified the bias on $r$ arising from correcting for the frequency-dependent polarization angle assuming a flat CMB spectrum on all frequency bands, and ignoring the presence of other components. In these circumstances, the $E$-$B$ rotation in synchrotron and dust caused by their different SEDs is completely ignored and could potentially cause a bias on $r$. However, we find that this bias is $\Delta r<10^{-4}$, and therefore negligible for SO.

One final potential concern for the polarization angle self-calibration strategy is the existence of intrinsic foreground $EB$ correlation \cite{2016MNRAS.457.1796A,2020PTEP.2020f3E01M}. With a non-zero foreground $EB$ component, the polarization angles can be mis-calibrated, which in turn leaks CMB $EE$ modes into $BB$ modes, biasing $r$. We assume the foreground $EB$ SED is the same as in $EE$ and $BB$, and parametrize the power spectrum with a power law with $\alpha^{EB} = (\alpha^{EE} + \alpha^{BB})/2$. The $EB$ amplitude is then varied between $\pm A^{BB}$. We find that the foreground-induced polarization angle biases are negligible for reasonable values of $A^{EB}$, for both dust and synchrotron. Biases of $10^{-3}$ are not possible with $A^{EB}<A^{BB}$. For dust and synchrotron, $A^{EB}/A^{BB}$ has to be greater than 75\% and 50\% respectively to cause a bias of $2\times10^{-4}$. These large amplitudes seem unlikely given the \planck constraint of $A^{EB}/A^{BB}<6\%$ on large patches of the sky~\citep{planck2018dust}. 

\subsection{Systematics marginalization} \label{ssec:results.std}

To reduce the biases from systematics errors in the bandpass and angle calibration, we include these parameters in the data model and marginalize over them. If the parametrization correctly characterizes the errors, and the marginalization prior contains the true value of the parameter, then this procedure will eliminate biases, potentially at the cost of reduced constraints on CMB parameters. We explore this marginalization cost in a variety of different scenarios and focus on the impact on the $r$ constraint. 

To produce a baseline for comparison we first estimate the posterior on the CMB and foreground parameters with no systematic errors. We then split the systematics into two categories: bandpass parameters and polarization angle parameters. We also consider increased foreground complexity through foreground decorrelation and intrinsic foreground $EB$ correlation. The results are summarized in Table~\ref{tab:marg}.

\subsubsection{Systematic-less}
We begin by measuring the constraints on $r$ without systematics but with several different CMB and foreground modeling assumptions. First, we have a baseline with input parameters $r=0$ and $r=0.01$. The resulting uncertainties in both cases are $\sigma_r=0.0014$ and $\sigma_r=0.0018$, respectively. Since some of the systematic parameters are multiplicative (e.g.,  gain uncertainties), they are potentially less relevant for a vanishing $r$, and therefore we use $r=0.01$ as our fiducial model. It is worth noting that, although we quote results in terms of the standard deviation $\sigma_r$, the posterior distributions are mildly asymmetric and non-Gaussian. The listed $\sigma_r$ is the mean of the $16^{th}$ and $84^{th}$ percentiles of the $r$ parameter posterior samples. Additionally, there will be some uncertainty in the parameter standard deviations given an MCMC chain of fixed length. We estimate this uncertainty on $\sigma_r$ by studying sub-chain parameter standard deviations, as described in Section~\ref{sec:methods}. The typical sub-chain variation on $\sigma_r$ is approximately $7\times10^{-5}$, up to $1\times10^{-4}$ for high dimensional models, and so differences in the constraints on $r$ must be understood in the context of these chain uncertainties. 

We tested the potential for improvement from delensing by using an input value of $A_{\rm lens}=0.5$, instead of 1, which gives $\sigma_r=0.0013$. For the same reasons above, we use $A_{\rm lens}=1$ as our fiducial model. We tested that extending the $\ell$ range from $\ell_{\rm max}=300$ to 600 as a way to constrain $A_{\rm lens}$ better does not provide a significant improvement on $\sigma_r$.

Foreground frequency decorrelation is one of the most important parameter extensions, given the rigid foreground modeling assumptions. We allow for decorrelation in both synchrotron and dust, and marginalize over $\pm 10\%$ top-hat priors \cite{2018PhRvD..97d3522S}. The data themselves constrain the decorrelation parameters for both foreground sources at the sub-percent level, although it increases the uncertainty on $r$ by $6\times10^{-4}$ (30\%) to $\sigma_r=0.0024$, as shown in Figure~\ref{fig:rpdf_bandpass}.

We also explored the additional information to be gained from including $EE$ information just by virtue of having more data with which to constrain the foreground parameters (we assumed the SED is the same in $E$ and $B$. Adding the $EE$ data to the model makes no difference on $r$ and $A_{\rm lens}$, although it does improve our estimates of the foreground spectral indices. If we assume that the decorrelation is also the same in $EE$ and $BB$, then we can actually improve our decorrelation measurement. This scenario brings $\sigma_r$ back down to $0.002$, an increase of only $2\times10^{-4}$ over the baseline (about 10\% degradation). It is worth noting that, in this case, we assume perfect knowledge of the CMB $EE$ power spectrum, and therefore this very mild improvement is likely overestimated. Studying the 2-D posteriors of these scenarios we find that $r$ and $A_{\rm lens}$ are only weakly correlated with any of the foreground parameters, except for the decorrelation parameters which are positively correlated with $r$. $r$ and $A_{\rm lens}$ are negatively correlated with each other. The values of $\sigma_r$ quoted here will be used a baseline for comparison when marginalizing over systematic parameters. 

\begin{table}[th!]
\centering
\begin{tabular}{| l | c | c |}
\hline
Case & $\sigma(r)$ & $\sigma(r)$, decorr. \\
\hline
\hline
\multicolumn{3}{|c|}{Systematic-less baseline} \\
\hline
$r=0.01$              & $1.8\times 10^{-3}$ & $2.4\times 10^{-3}$\\
$r=0$                 & $1.4\times 10^{-3}$ & \\
$EE$ and $EB_{\rm FG}$         & $1.8\times 10^{-3}$ & $2.0\times 10^{-3}$\\
\hline
\multicolumn{3}{|c|}{Bandpass uncertainties} \\
\hline
$\Delta\nu_i$, $\Delta g_i$ (3\% top-hat prior) & $1.9\times 10^{-3}$ & $2.6\times 10^{-3}$\\
$\Delta\nu_i$, $\Delta g_i$ with $EE$ and $EB_{\rm FG}$   & $1.8\times 10^{-3}$ & $2.1\times 10^{-3}$\\
$\Delta\nu_i$, $\Delta g_i$ (1\% Gaussian prior)      & $1.9\times 10^{-3}$ & \\
$\Delta\nu_i$, $\Delta g_i$ (3\% Gaussian prior)      & $2.0\times 10^{-3}$ & \\
$\Delta\nu_i$, $\Delta g_i$ (5\% Gaussian prior)      & $2.2\times 10^{-3}$ & \\
\hline
\multicolumn{3}{|c|}{Polarization angle uncertainties} \\
\hline
$\Delta\phi_0$, $\Delta\phi_1$, $EE$,              & $1.8\times 10^{-3}$ & \\
$\Delta\phi_0$, $\Delta\phi_1$, $EE$, $EB_{\rm FG}$ & $1.8\times 10^{-3}$ & $2.0\times 10^{-3}$\\
\hline
\multicolumn{3}{|c|}{Bandpass and angle uncertainties} \\
\hline
$\Delta\nu_i$, $\Delta g_i$, $\Delta\phi_0$, $\Delta\phi_1$, $EE$, $EB_{\rm FG}$ (3\% top-hat prior) & $1.8\times 10^{-3}$ & $2.1\times 10^{-3}$ \\
\hline
\end{tabular}
\caption{Uncertainty on $r$ showing the impact of different modeling assumptions and systematic parameter marginalization. The systematic-less runs show that the foreground decorrelation parameter strongly effects the ability to constrain $r$, increasing $\sigma_r$ by 33\%. Assuming the foreground SEDs are the same in $E$ and $B$ allows us to reduce the degradation from decorrelation to about 10\%. Marginalizing over bandpass uncertainties only weakly degrades the $r$ constraint for typical prior widths.  We note in parentheses the priors used for the bandpass parameters, which are mostly unconstrained by the data. Marginalizing over polarization angles produces negligible changes to the $r$ constraint, as the angles are self-calibrated using the CMB $EB=0$. We used sub-chain variations to estimate that these constraints are accurate to $\approx7\times10^{-5}$ (e.g.,  a change of $1\times10^{-4}$ in $\sigma_r$ is a only 1-$\sigma$ difference, considering the uncertainties on both runs). The SO forecast for $\sigma_r$ lies in the range of $2-3\times10^{-3}$ depending on the specifics of the foreground model, noise, and component separation.}
\label{tab:marg}
\end{table}

\subsubsection{Bandpass uncertainties}
We assess the impact of bandpass calibration errors in the form of gain calibration and frequency shift parameters. The gains and frequency shifts only affect the frequency dependence of the various sky signals and do not change their scale dependence in any way. The result is that marginalizing over bandpass parameters primarily impacts the foreground spectral indices and amplitudes, which in turn can leak into the CMB $r$ and $A_{\rm lens}$ amplitudes. Since the CMB blackbody spectrum is well known, and given the significant amount of $\ell$-domain information, the foreground SED modifications only weakly impact the CMB amplitudes, as we see in a variety of different scenarios below and listed in Table~\ref{tab:marg}. 

\begin{figure}[tbp]
\centering
\includegraphics[width=0.54\textwidth]{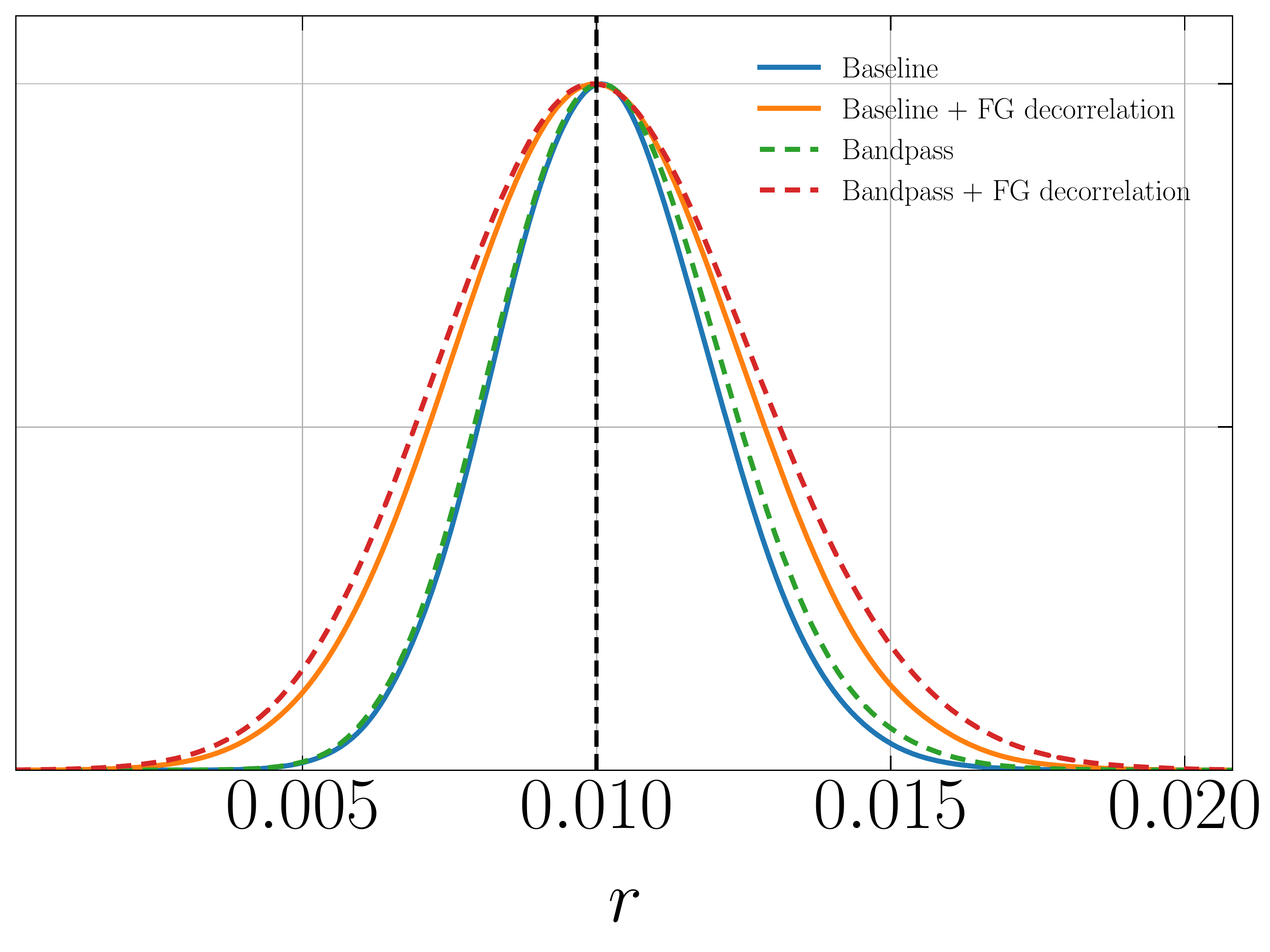} \hfill
\includegraphics[width=0.44\textwidth]{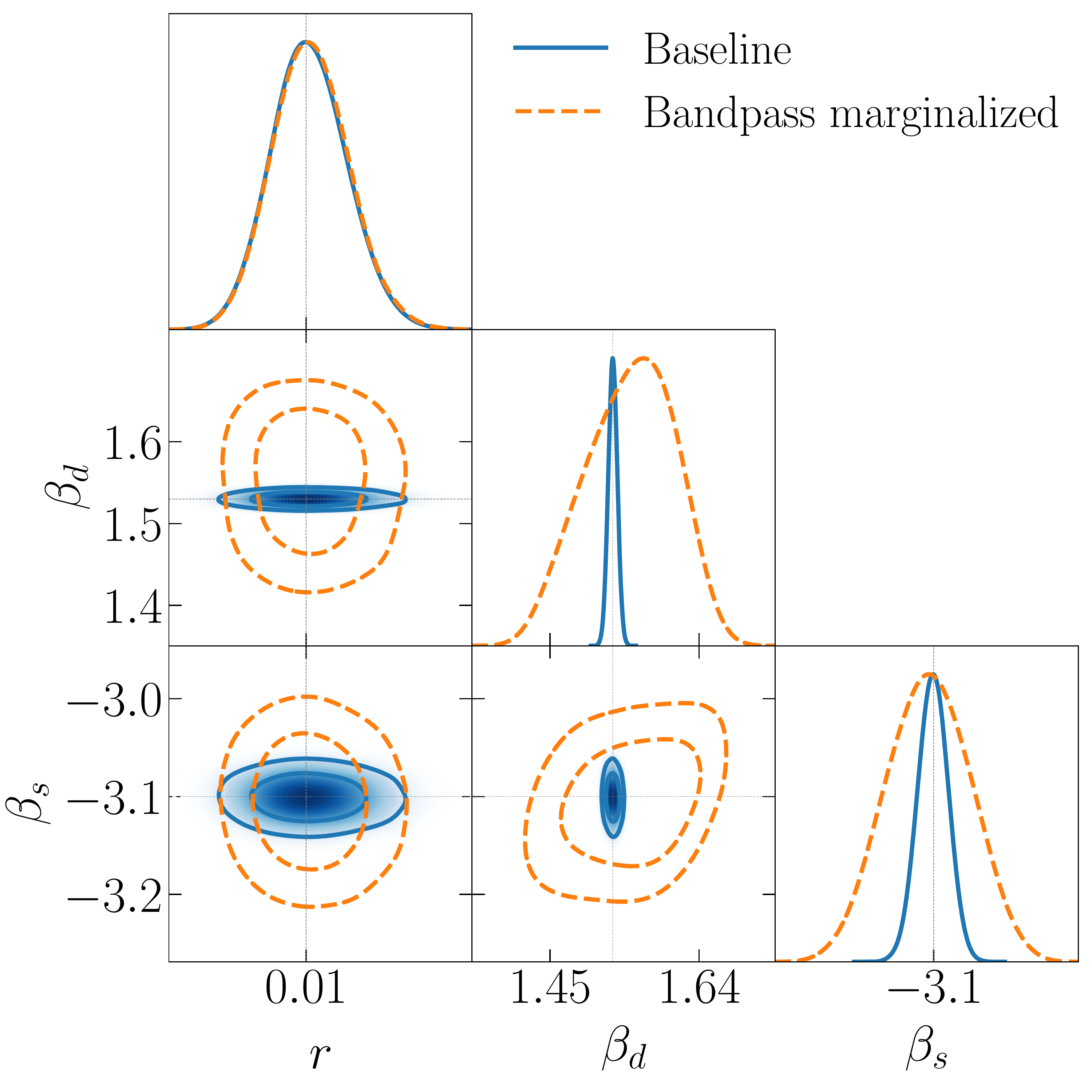}
\caption{(Left) Marginalized posterior on $r$ comparing bandpass uncertainties and foreground decorrelation. The systematic-less models are in solid lines and the bandpass marginalized are in dashed lines. The bandpass marginalization alone does not significantly hamper the constraint on $r$ (comparing solid blue and dashed green). Including foreground decorrelation (solid orange) shows a modest reduction in the constraint on $r$, from $\sigma_r=0.0018$ to $\sigma_r=0.0024$ (without $EE$ information). The bandpass marginalized with foreground decorrelation is the worst-case with $\sigma_r=0.0026$, approximately 40\% larger than the baseline. (Right) Posterior of $r$ and the foreground spectral indices $\beta_d$ and $\beta_s$ for the systematic-less baseline and bandpass marginalized cases. The posterior on $r$ is hardly impacted by the bandpass parameter marginalization, while the foreground spectral indices have significantly broadened constraints. }
\label{fig:rpdf_bandpass}
\end{figure}

Marginalizing over $\pm 3\%$ top-hat uncertainties in gains and frequency shifts with $BB$ data only produces $\sigma_r=0.0019$, which is only a $1\times10^{-4}$ increase in $\sigma_r$ (5.6\% increase), as seen in Figure~\ref{fig:rpdf_bandpass}. Most of the uncertainty from this marginalization is absorbed into $A_{\rm lens}$ which has its uncertainty increased by a factor of 2 from from 0.025 to 0.052, since lensing B-modes dominate the CMB amplitude. The $\pm 3\%$ prior should be wide enough to account for reasonable calibration errors of the bandpasses. After calibrating the final spectra to the \planck and \wmap spectra, current small aperture experiments typically achieve $<\sim5$\% gain calibration uncertainties~\citep{ ABS2018, QUIET_2011,2019ApJ...876..126A}, while calibration uncertainties are on order $\sim1$\% for arcminute-resolution experiments~\citep{2020JCAP...12..045C,act2013calibration, planck2017planetflux,2018ApJS..239...10C}. We also explored Gaussian priors with 1, 3, and 5\% widths. The increased prior width corresponds to more and more degradation of the $r$ constraint to $\sigma_r=0.0019$, $0.0020$, and $0.0022$, respectively (an increase of roughly 6\%, 11\%, and 22\% over baseline).

For appropriately sized bandpass uncertainty priors, the marginalization cost on $r$ is fairly small. However, additional foreground parameters might interact poorly with these systematics. We therefore consider the case of foreground decorrelation. With the 3\% top-hat bandpass priors and 10\% top-hat decorrelation priors we get $\sigma_r=0.0026$ an increase of 8\% over the systematics-less baseline that also includes decorrelation ($\sigma_r=0.0024$). Roughly, the decorrelation costs an additional $6\times10^{-4}$ in $\sigma_r$, the bandpass marginalization incurs a $1\times10^{-4}$ penalty, and the bandpass-decorrelation cross term also contributed $1\times10^{-4}$ to $\sigma_r$.

Interestingly, the rigidity of the CMB SED and foreground model allows for self-calibration of the bandpass parameters to some extent, in particular, the MF and UHF frequency shift parameters. This is due to the sensitivity of the model to these shifts. As seen from the bias requirement study, the MF and UHF shifts have the tightest bounds, and therefore large departures in these parameters are easily detected in the residuals. With the assumption that the gains and shifts are the same in $EE$ and $BB$, adding $EE$ information significantly improves the self-calibration of the bandpass parameters due to the brightness of the CMB $E$-modes\footnote{It is worth noting again that we have assumed a perfectly-known $E$-mode CMB spectrum.}.

The bandpass parameters do not appear to be very degenerate with $r$, as shown in Figure~\ref{fig:rpdf_bandpass}. We saw in the bias study in Section~\ref{ssec:results.bias} that only a specific combination of bandpass parameters are degenerate with $r$ in a way that is not detected by an obvious increase in the $\chi^2$. For example, we already found that symmetric gain and shift combinations produce significantly less bias on $r$ than anti-symmetric combinations as the coherent variations can be absorbed almost completely into foreground amplitudes. Systematic contributions that produce strong deviations away from the data can thus be detected through any simple goodness of fit test. Therefore, although the systematics show many degeneracies and are not well constrained, the overall volume of systematic parameter space that affects $r$, while simultaneously not contributing significant residuals to other components of the model, is small.

Therefore, given an accurate bandpass error and foreground model, we can marginalize over gains and shifts without a significant impact on our ability to constrain $r$. The limiting factor is not the volume of the systematics prior space but rather the foregrounds and instrument noise unless we can delens to the extent that the primordial B-modes becomes a dominant amplitude in the model. Mismodeling of the foregrounds could complicate things, but the foreground decorrelation results suggest that any foreground-bandpass interaction is completely subdominant to the constraint degradation from the additional foreground complexity itself.

\subsubsection{Polarization angle uncertainties}

The polarization angle systematics affect the scale dependence of the model, but not the frequency spectrum, since we assume the $E$ and $B$ modes of all components have the same frequency dependence. However, using the vanishing CMB $EB$ correlation allows us to measure the angles and angle tilts such that there is no impact on $r$ from this marginalization, and $\sigma_r=0.0018$.
As one of the key assumptions, we test the impact of marginalizing over intrinsic foreground $EB$, which could bias the CMB $EB$ self-calibrated angles~\cite{2016MNRAS.457.1796A}. Data from \planck so far have constrained dust $EB$ to be consistent with zero and less than 3\% of the $EE$ amplitude (approximately 6\% of $BB$). We allow for a power-law intrinsic foreground $EB$ in both synchrotron and dust and marginalize over $EB$ amplitudes and power spectrum indicies, with the SED index fixed to that of $EE$ and $BB$. This reduces our constraining power on the angles, but still is not significant enough to impact the uncertainty on $r$.

\subsubsection{Bandpass and angle uncertainties}
The combination of both bandpass and angle uncertainties is dominated by the bandpass contribution, as the angles are well determined from the $EB=0$ assumption. After marginalizing over 43 parameters, including intrinsic foreground $EB$ and foreground decorrelation, we find no statistically significant change in $\sigma_r$ given the chain uncertainties. We had difficulty with convergence for the combined bandpass and angle runs. The Gelman-Rubin test failed for several parameters. However, the chains are 10000 times the auto-correlation length (order $10^8$ samples total) and the Gelman-Rubin test was satisfied for $r$ so we terminated the runs. In the future we will need to consider more efficient sampling methods.

\subsection{The \bk data}\label{sec:bicep}

To test the validity of the main results of this analysis on a more realistic setup, we have repeated our analysis on the latest public data from the \bk collaboration, presented in \cite{2018PhRvL.121v1301B} (BK15X henceforth). The aim of this exercise is threefold: first, it allows us to test our analysis pipeline, verifying that it is able to accurately model realistic levels of foreground contamination. Secondly, we can test the validity of our results on noisy data (recall that our analysis so far has been carried out on mock power spectra with no statistical noise). Finally, doing so allows us to quantify the robustness of the published BK15X results to the presence of residual bandpass/angle systematics. 

A complete description of the BK15X dataset can be found in \cite{2018PhRvL.121v1301B} and references therein. The data combines information from 12 different frequency bands, including the three \bk bands (at 95, 150 and 220 GHz), 2 low-frequency \wmap bands (23 and 33 GHz) \cite{2013ApJS..208...20B} and 7 \planck bands (30, 44, 70, 100, 143, 217 and 353 GHz) \cite{2016A&A...594A...1P}. Power spectra were computed over the 400 deg$^2$ \bk patch, and all cross-correlations between different frequency and polarization bands are made publicly available together with their covariance matrix as well as all frequency bandpass transmission curves and bandpower window functions\footnote{See \url{http://bicepkeck.org/bk15_2018_release.html}.}. Power spectra are computed in 9 bandpowers covering the multipole range $30\leq\ell\lesssim400$. The size of the complete data vector, including all $BB$, $EE$ and $EB$ cross-correlations is 2700 elements, which reduces to 702 if only $B$-mode data is used.

\begin{figure}[tbp]
\centering
\includegraphics[width=0.48\textwidth]{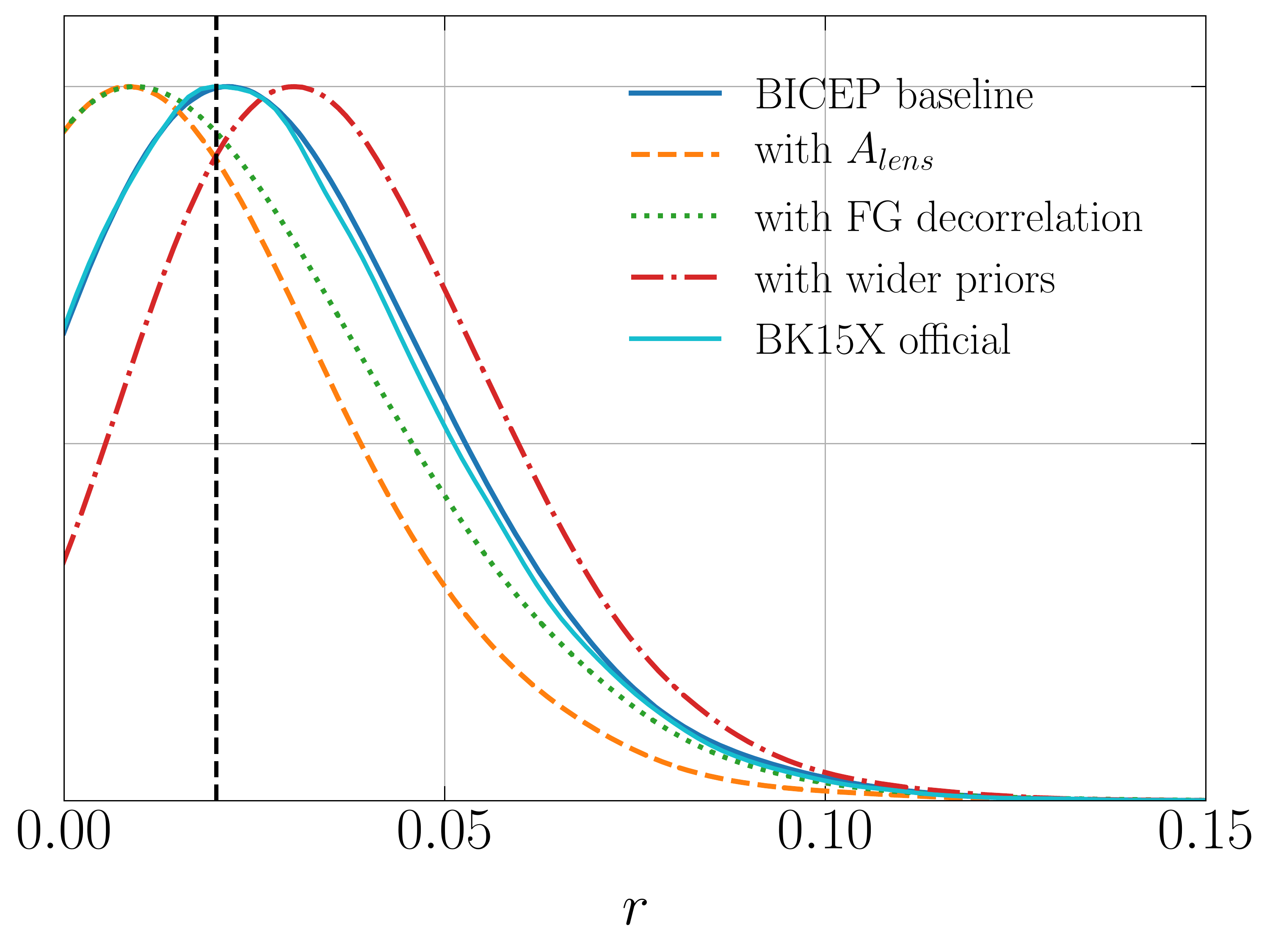} \hfill
\includegraphics[width=0.48\textwidth]{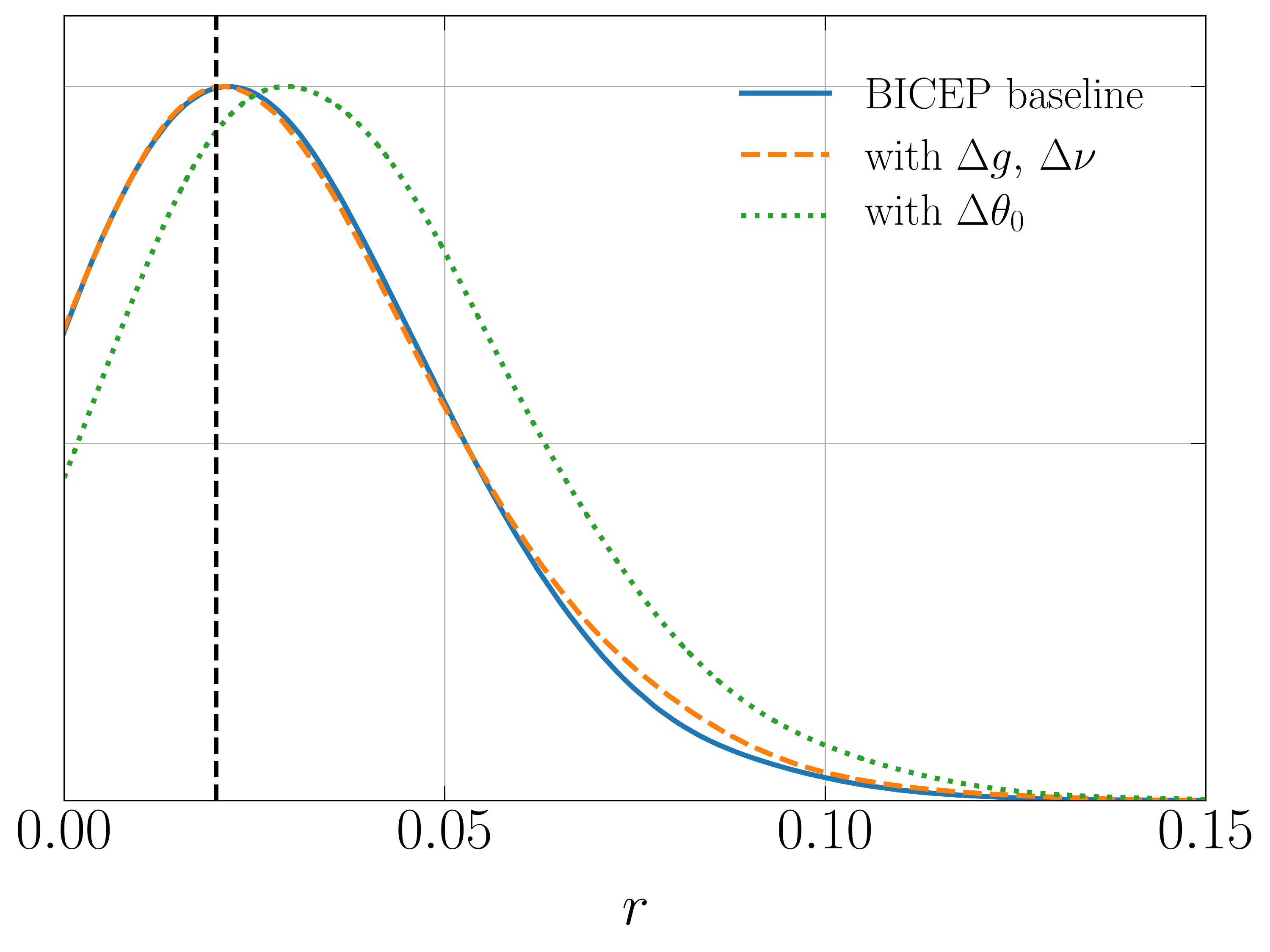}
\caption{(Left) Marginalized posterior on $r$ using our pipeline on the publicly available BK15X data. The blue line shows the baseline BK15X model and agrees well with their published tabulated likelihood shown in light blue. The vertical black dashed line corresponds to r=0.02, the peak of the BK15X posterior. We test several scenarios that include parameters beyond the baseline ones, similar to the modifications in B15X. Marginalizing over $A_{\rm lens}$ shifts the curve closer to zero (orange). Likewise when allowing for dust decorrelation (green), the curve moves towards zero. Interestingly, increasing the width of the priors on the power law indices $\alpha$ from $[-1, 0]$ to $[-4, +1]$ (red) pushes for a larger value of $r$. (Right) Posterior on $r$ from the BK15X data, marginalizing over bandpass and angle systematics. The baseline systematics-less case is in blue. Marginalizing over 3\% top-hats for gains and frequency shifts for all 12 bands does not significantly impact the $r$ posterior (orange). The angle marginalized posterior (green) moves towards slightly larger values of $r$.}
\label{fig:bicep_syst}
\end{figure}

We first attempt to reproduce the fiducial BK15X results by running our components separation pipeline on the $B$-mode data with all systematic parameters fixed to zero, and all parameter priors fixed to the same values used by BK15X. The results are displayed in Figure~\ref{fig:bicep_syst}, which shows our recovered posterior distribution on $r$ in solid blue, together with the publicly available tabulated likelihood found by BK15X in light blue. We are able to recover the posterior distribution well and parameter constraints agree to better than 5\%. We also find visual agreement between our multi-dimensional contours involving all other foreground parameters and those presented in BK15X. In order to test the validity of the foreground model implemented in our pipeline, we have also explored extensions of the fiducial setup. These are shown in Figure~\ref{fig:bicep_syst}, and involve freeing up the $B$-mode lensing amplitude (orange), allowing for non-zero frequency decorrelation in dust (green), and broadening the foreground parameter priors (red). The dust decorrelation prior is a 10\% top-hat as in our SO analysis. We increased the foreground power spectral index priors from [-1, 0] to [-4, 1] to match our SO analysis priors. We observe that the resulting posteriors shift and broaden slightly, especially when allowing for decorrelation, as reported in BK15X.

We then explored the impact of bandpass/angle systematics on the BK15X data. While the BK15X data come from several instruments with designs that are different from the SO SAT, bandpass and average polarization angle systematic uncertainties are common to all experiments. We first allowed for changes in the center frequencies and gains of the 12 frequency bands, corresponding to 24 new parameters ($\Delta\nu_b$, $\Delta g_b$, $b\in[1,12]$), on which we impose a $3\%$ Gaussian prior, motivated by the \wmap calibration uncertainties. The results, in this case analysing only $BB$ data, are shown in orange in Figure~\ref{fig:bicep_syst}, together with our fiducial results. In agreement with our fiducial results, the posterior distribution on $r$ is relatively insensitive to these systematics, and the penalty for these bandpass uncertainties is absorbed by the foreground parameters. The 95\% upper bound on $r$ is 0.077 for the baseline systematic-less case and 0.081 with gain and shift marginalization, a 5\% increase in $\sigma_r$.

We have also quantified the impact of polarization angle systematics by allowing for a constant angle parameter $\Delta\phi_{0,b}$ to vary in all frequency bands with a $\pm10^\circ$ prior. In this case we also included all cross-spectra involving both $E$ and $B$ modes. The results are shown in green in Figure~\ref{fig:bicep_syst}. We observe a small shift on the posterior distribution. This is caused by a small shift away from zero of some of the polarization angles, although all of them are compatible with zero, which shifts some $E$-mode power into $B$-modes. The width of the posterior however remains virtually unchanged, with $\sigma_r=0.026$, in agreement with the results presented in the previous section. It is worth emphasizing that the formalism used to obtain these results implicitly assumes a null CMB $EB$ correlation, which is a potentially strong theoretical prior.

In conclusion, we find that the main result presented in Section~\ref{ssec:results.std}, i.e.,  that marginalizing over bandpass and angle uncertainties within reasonable priors has a negligible effect on the constraints on $r$, also holds when our analysis pipeline is applied to the real BK15X data. 

\begin{figure}[tbp]
\centering
\includegraphics[width=0.6\textwidth]{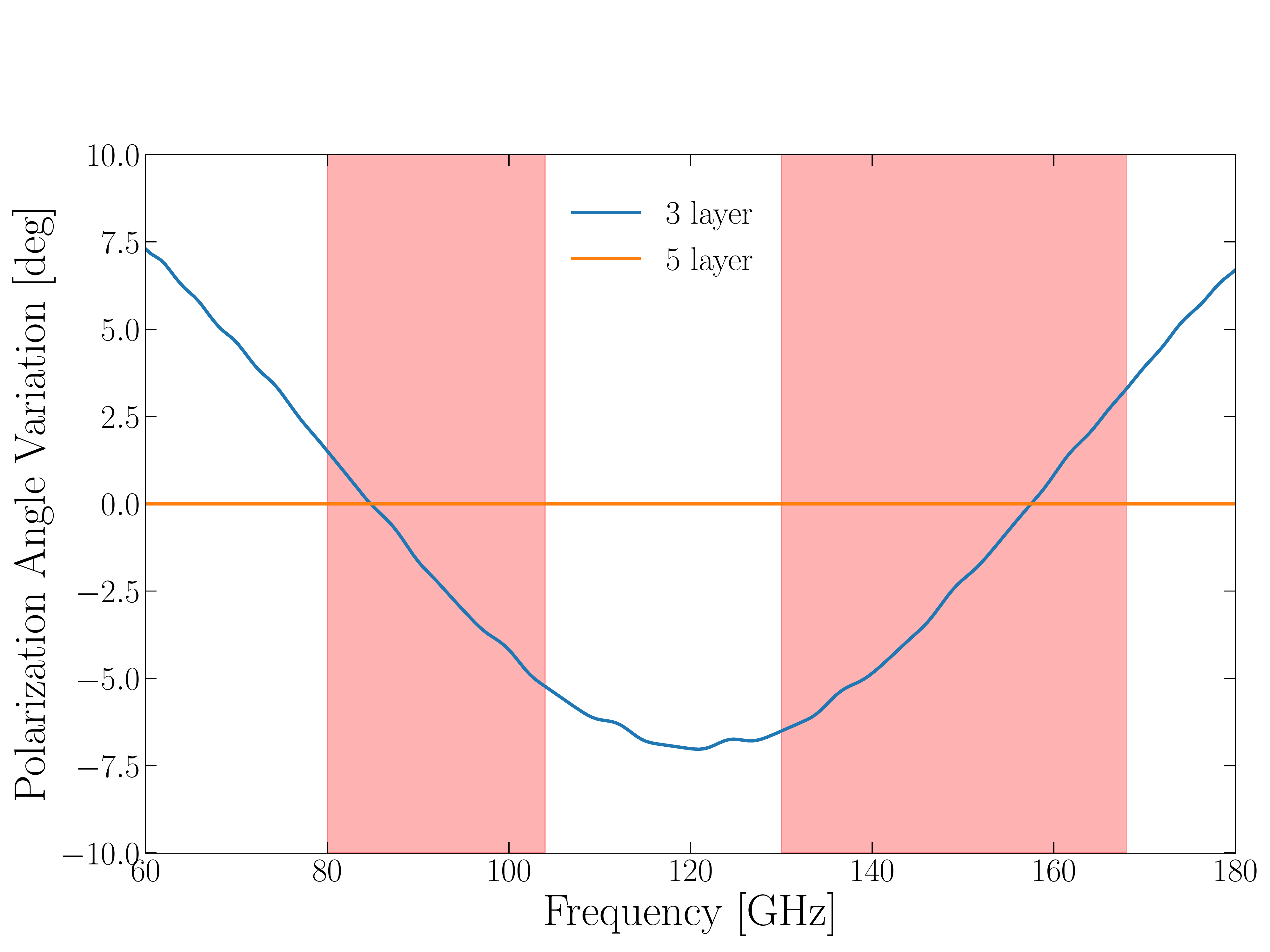}
\caption{Polarization angle variation as a function of frequency for example 3-layer and 5-layer HWPs with a three-layer AR coating. The shaded red regions show the SO MF bands. The 5-layer design eliminates essentially all polarization angle variation, but is more difficult to implement in the SO SAT system.
}
\label{fig:HWP}
\end{figure}

\section{Implications for instrument design and calibration}\label{sec:design}
The requirements on the bandpass and polarization angles in Table~\ref{tab:syst_req_table} have several implications both for the instrument design and the calibration plan for SO and future CMB experiments. In this section, we discuss how these requirements propagate to design and calibration decisions for the SO SATs.

\subsection{Implications for instrument design}
In the following section, we discuss the implications of the requirements derived in Table~\ref{tab:syst_req_table} on two components of the instrument design: the HWP optical design and the detector time constants. Because the HWP is modulating the polarization angle, the requirements from Table~\ref{tab:syst_req_table} set requirements on the HWP optical design. The phase of the polarization modulation from the HWP determines the polarization angle of the detectors. However, the detector time constants introduce a phase lag, which can cause uncertainty in the polarization angle. The polarization angle requirements thus place constraints on the detector time constants and their uncertainties as well~\cite{Time_const_2014}.

\subsubsection{SAT half-wave plate design}\label{sssec:css.design.hwp}

The SAT HWP uses sapphire slabs to create a net retardation of $\lambda / 2$ between two orthogonal polarizations. The retardation of the sapphire varies as a function of frequency which reduces the polarization modulation efficiency and varies the polarization angle away from the center frequency. By stacking more layers of sapphire with the correctly tuned thicknesses and rotation angles, one can produce an achromatic HWP that has close to 100\% modulation efficiency and reduced polarization angle variation across a wider range of frequencies \cite{Matsumura806analysisof}. The baseline HWP design of the SAT was a 3-layer sapphire HWP, but the allowable polarization angle variation could have necessitated a HWP design with more layers. A design with more layers would have added both additional cost and mechanical risk as a HWP with more layers would be thicker and heavier, pushing the limits of the spatial and mechanical constraints within the SAT optics tube.

\begin{table}[h!]
\centering
\begin{tabular}{|c|c|c|c|c|}

\hline
& \multicolumn{2}{c}{$\Delta \phi$ sync (deg)} & \multicolumn{2}{|c|}{$\Delta \phi$ dust (deg)}\\
\hline
 shift (\%) & 94 GHz & 148 GHz &  94 GHz & 148 GHz \\
\hline
 0 &          0.000 &           0.000 &          0.000 &           0.000 \\
 1   &          0.029 &           0.012 &          0.023 &           0.012 \\
 3   &          0.087 &           0.025 &          0.067 &           0.025 \\
 5   &          0.147 &           0.028 &          0.105 &           0.028 \\
\hline
\end{tabular}
\caption{
The variation in the band-averaged polarization angle of a 3-layer HWP due to bandpass shift for the synchrotron and dust components is shown above, assuming perfect CMB calibration. Spectral indices used are $\beta_s = -3$ and $\beta_d = 1.59$ for the synchrotron and dust components respectively. All shifts are less than the most stringent 0.2$^\circ$ requirement. } 
\label{tab:hwp_spec_angles}
\end{table}

To come to this decision, we simulated 3, 4, and 5 layer HWPs and compared their polarization modulation efficiencies and polarization angle variations across the 94/148 GHz band since the polarization requirement is most stringent for these bands.
To simulate the HWPs we used a generic transfer matrix method \cite{essinger-hilemanSystematicEffectsAmbienttemperature2016} to generate the HWP Mueller Matrices as a function of frequency and rotation angle $\chi$.
We propagate polarized light through the HWP Mueller Matrix to obtain the modulated signal as seen by the detector, and fit this signal to the sum of the first eight Fourier modes \cite{kusakaModulationCMBPolarization2014}. The modulation efficiency and polarization angle are then estimated from the $n=4$ harmonic amplitude and phase respectively \cite{kusakaModulationCMBPolarization2014}.

Optimal thicknesses and rotation angles of each layer were determined through basin-hopping optimization \cite{matsumuraPrototypeDesignEvaluation2018}.
We determined that for the correct thicknesses and rotations, 3 and 5 layer HWPs produced good polarization modulation profiles, and 4 and 5 layer HWPs could essentially eliminate polarization angle variation across the 94/148~GHz bandpasses as can be seen in Figure~\ref{fig:HWP}. We thus considered both the 3-layer and 5-layer designs. 

As discussed in Section~\ref{sec:methods}, if the polarization angle variation and bandpasses are well-known as a function of frequency, they can be included in the data model to remove any biases on $r$. In practice, there will be uncertainties in the polarization angle variation function, in addition to the bandpass uncertainties which were discussed previously. One of the primary issues with the frequency-dependent polarization angle rotation is it effectively assigns different band-averaged angles to signals with different SED frequency dependence. The instrument therefore sees rotated versions of the polarized CMB, synchrotron, and dust skies, but each with a different rotation, and the rotation also changes band-to-band. 

We can characterize uncertainties in the band-averaged rotations for each signal by considering the impact of bandpass shifts on the integration. We calculate how much a bandpass shift of up to 4~GHz changes the band-averaged polarization angle for the CMB, dust, and synchrotron components individually. We find that for the 3-layer HWP, if we assume perfect calibration in the CMB band, the polarization angle for the dust and synchrotron components will shift by at most $0.15^\circ$ as shown in Table~\ref{tab:hwp_spec_angles}. This is within the most stringent polarization angle requirements in Table~\ref{tab:syst_req_table}, which were more generally defined assuming all signals were rotated by the same amount. 

\begin{figure}[t!]
\centering
\includegraphics[width=0.5\textwidth]{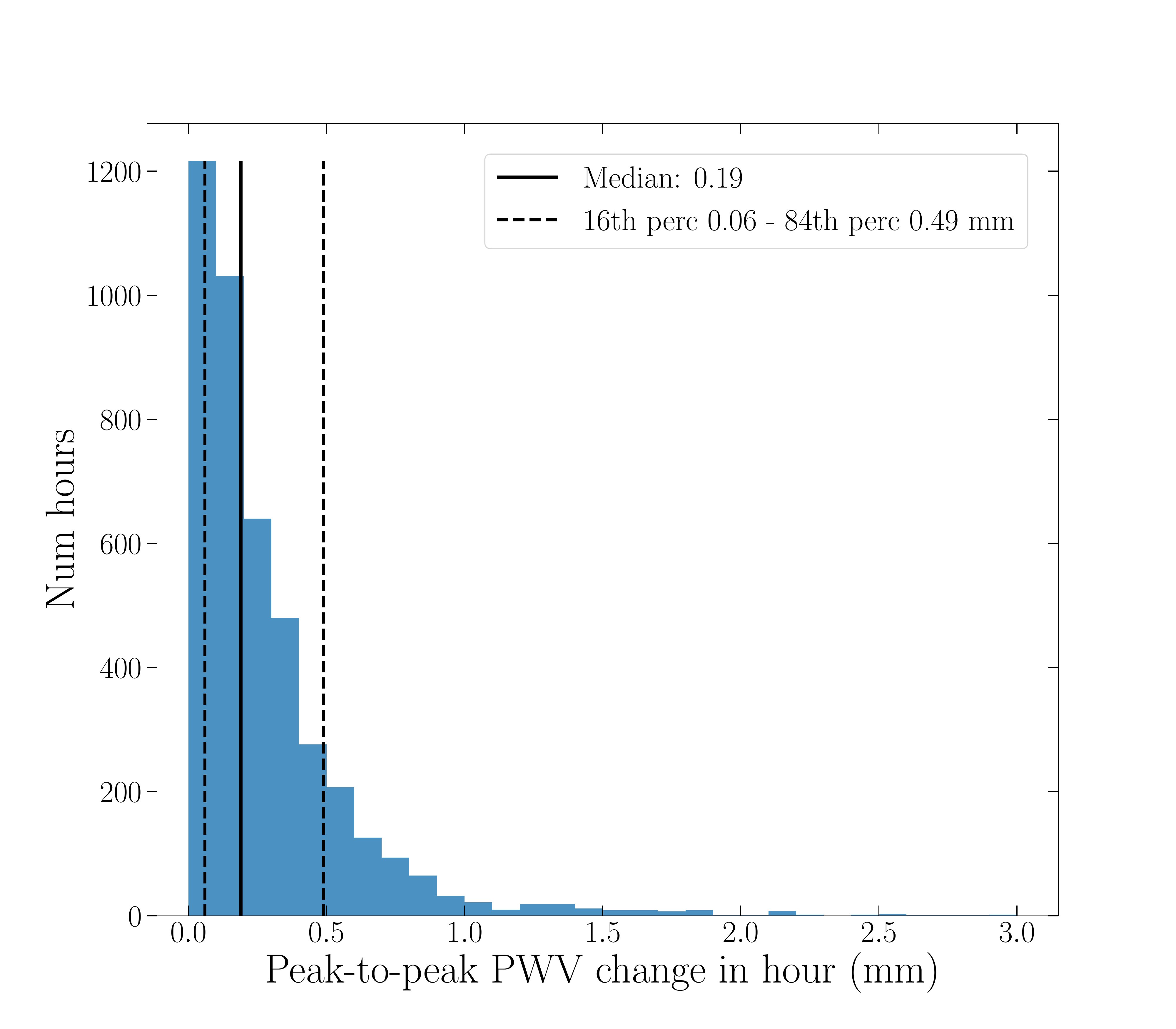}
\caption{Histogram of peak-to-peak PWV values for hours with initial PWV between 0~mm and 3~mm. We set 0.5~mm as a conservative value for excursions of the PWV value during observations.The solid vertical line is the median of the distribution, while the left and right dashed lines indicate the 16th and 84th percentile, respectively.}
\label{fig:pwv_data}
\end{figure}

The worst-case would be to ignore the frequency-dependent rotation completely, which leads to a bias of $1.5\times10^{-4}$. Within a given band, we find that the frequency-dependent angle variation function is well-described by a first-order expansion, as introduced in Section~\ref{sec:methods}. Generating data with the simulated HWP polarization angle function but fitting with a model consisting instead of $\Delta\phi_0$ and $\Delta\phi_1$ completely eliminates the bias on $r$ with only a marginal cost in $\sigma_r$, similar to the sinuous antennas results in Figure~\ref{fig:sinuous}. This, together with the results of the likelihood-level study presented in Section \ref{ssec:results.bias} led us to select the 3-layer HWP design for the SAT.

\subsubsection{Detector time constants}
When operating with a HWP polarization modulator, the apparent polarization angle of a given detector is determined both by the intrinsic detector angle and the time constant response of the detector. Both values enter the expression for the phase of the modulated polarization signal at frequency \fsig $= 4$ \frot, where \frot is the rotation frequency of the HWP~\cite{Time_const_2014}.

The SO detectors are transition-edge sensor bolometers, and we approximate their temporal response as a single-pole low-pass filter with \fdb = $1/(2 \pi \tau_{\mbox{\scriptsize eff}})$, where $\tau_{\mbox{\scriptsize eff}}$ is the time constant of the bolometer with passive negative feedback~\citep{irwinhilton2005}.

The apparent detector angle $\psi'$ is two times the phase of the detector response to the signal at \fsig and can be expressed as:

\begin{equation}
\psi' = \psi + \frac{1}{2} \arctan \left( \frac{f_{\mbox{\scriptsize sig}}}{f_{\mbox{\scriptsize 3dB}}} \right).
\label{psi_prime_def}
\end{equation}

\begin{figure}[t!]
\centering
\includegraphics[width=0.48\textwidth]{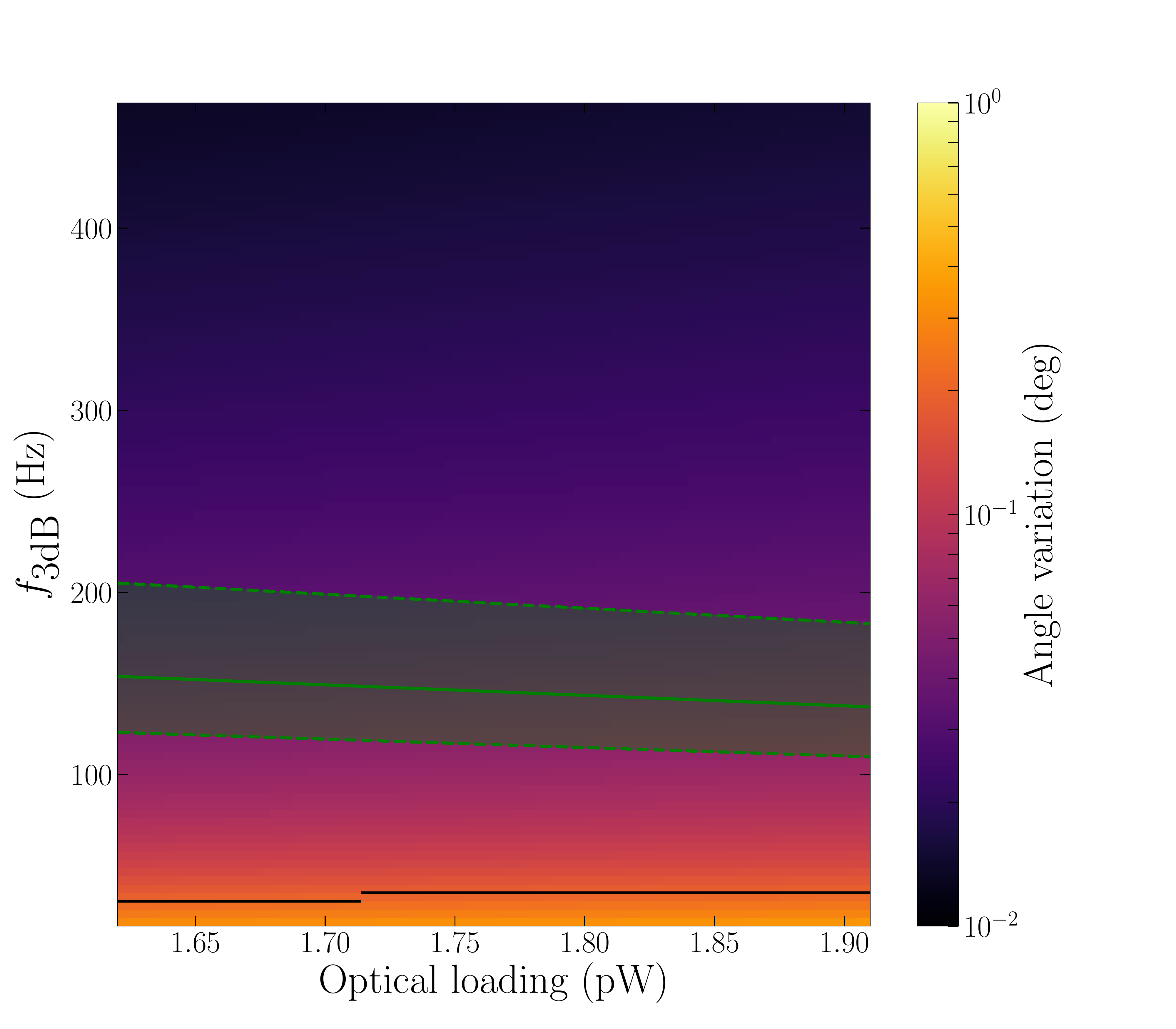}
\includegraphics[width=0.48\textwidth]{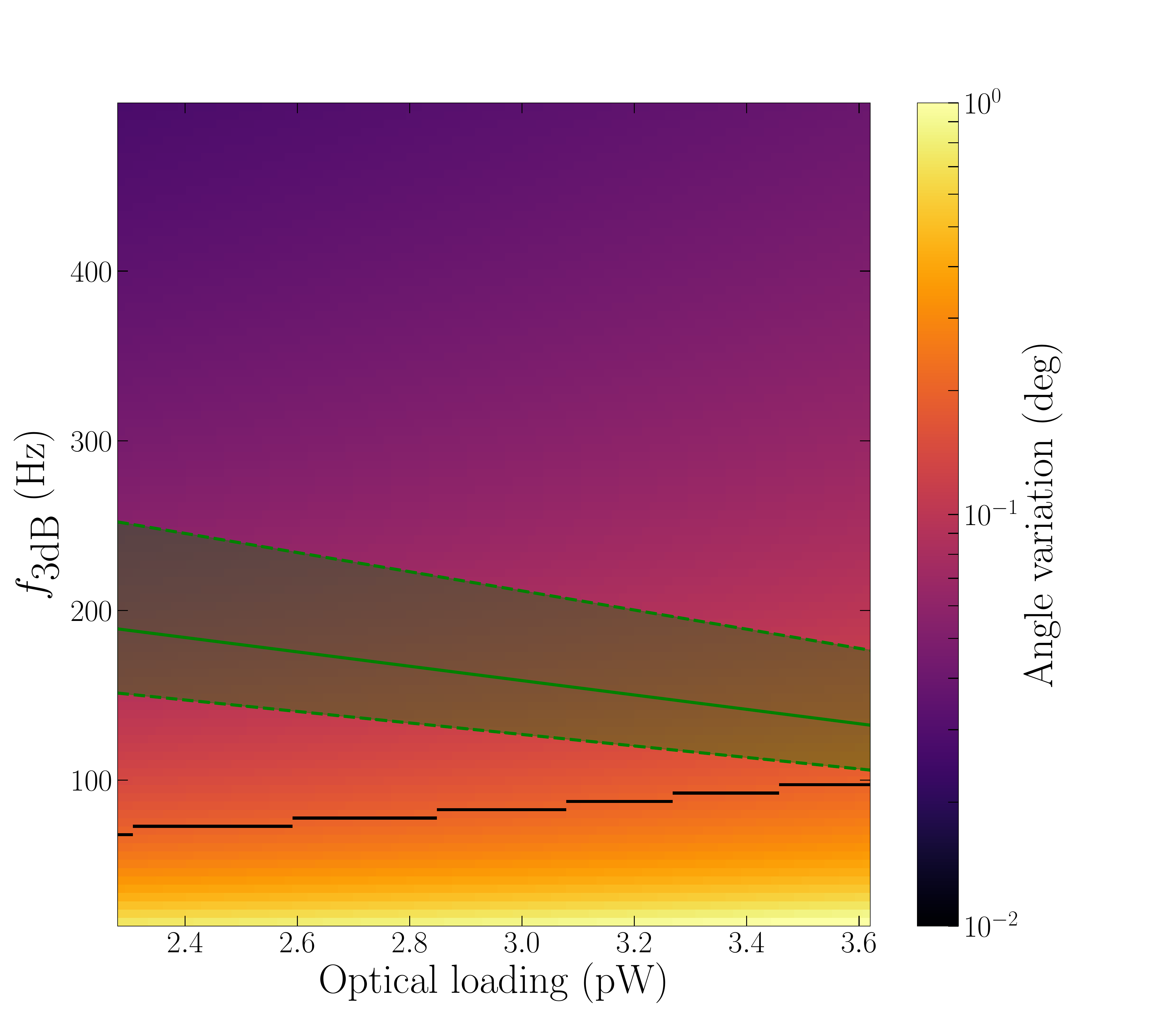}
\caption{Absolute value of expected polarization angle shift (color scale) for a fixed amount of PWV and optical power drift and fixed \fsig $= 8$~Hz for SO detectors in the 94~GHz (left) and 148~GHz (right) observing bands. The x-axis spans the expected optical loading on the detectors for static PWV values ranging from 0.5~mm to 2.5~mm. The y-axis represents achieved detector time constants for a wide range of possible detector designs. We expect SO detectors to fall within the green band representing achievable \fdb values. Summary statistics of the angle shifts within the green bands can be found in Table~\ref{tab:angle_shift_spec}. The black line indicates the allowable polarization angle uncertainty from Table~\ref{tab:syst_req_table}.}
\label{fig:angle_shift_map_MF}
\end{figure}

To first order, the contribution from the time constant can be calibrated and removed; however, variations in the time constant appear as apparent polarization angle shifts, resulting in systematic uncertainty in the polarization angle.
This effect is largely due to the detector \fdb fluctuating with loading changes. During SO SAT observations over the course of $\sim$1~hr, detector gain and temporal responses drift due to variation in the atmospheric brightness. We calibrate the electrical time constants of the detectors between each $\sim$hourly observation using bias steps and current versus voltage curves~\cite{Dunner_ACT_2012}~\cite{Grace_ACTPol_time_constants}. Periodic optical time constant measurements using a wire grid and varying HWP speed are used to correlate the electrical time constants to optical time constants~\cite{Simon_2014}. While there is time variation of the time constant throughout the observation, we assume a fixed time constant during the observation in the analysis. However, we can model changes to \fdb within each hourly observation due to loading and calculate the corresponding change in polarization angle. This change in \fdb, here labeled $\Delta f_{\mbox{\scriptsize 3dB}}$, is estimated directly from the expression for \fdb in terms of bolometer bias power in Eq. 29 of~\cite{irwinhilton2005} given that a change to detector loading is, in the simplest model, equal and opposite to a change in the bias power. Our estimation of the change to the polarization angle is then:

\begin{equation}
\Delta \psi' = \frac{1}{2} \left( \arctan \left( \frac{f_{\mbox{\scriptsize sig}}}{f_{\mbox{\scriptsize 3dB}} + \Delta f_{\mbox{\scriptsize 3dB}}} \right) - \arctan \left( \frac{f_{\mbox{\scriptsize sig}}}{f_{\mbox{\scriptsize 3dB}}} \right) \right).
\label{delta_psi_deg}
\end{equation}

The size of this effect depends on the expected static optical power incident on the detectors from the sky and instrument and a value for the expected drift of the optical power due to changes in the atmosphere. The atmospheric change most relevant for SO is a change to the precipitable water vapor (PWV). Fluctuations in PWV change the atmospheric transmission and absorption, which in turn changes the loading on the detectors. To determine relevant values for the change in PWV, labeled $\Delta \mbox{PWV}$, over hour timescales at the telescope site, we use radiometer measurements from the Atacama Pathfinder Experiment (APEX) team, a telescope located near the SO observation site on the Chajnantor Plateau in the Atacama Desert~\citep{2006_APEX, APEX_operations_2007}. Figure~\ref{fig:pwv_data} shows a histogram of the peak-to-peak PWV value across 1-hour units of time for hours in the date range May 2016 to Jan 2017, or about 1 season of observing time. These are hours when the first PWV value for the hour is less than 3~mm, a rough cutoff for observations being too insensitive for use in CMB science results. This histogram shows that 0.5~mm, rounded up from the 84th percentile of 0.49~mm in the figure, is a reasonable upper bound on the maximum hourly PWV drift expected.

With this value set, we estimate the apparent angle shift $\Delta \psi'$ generated by these PWV drifts and determine acceptable regions of detector parameter space that would allow the SO SAT observations to neglect uncalibrated drifts. We indicate these for the 94 and 148~GHz detector bands in Figure~\ref{fig:angle_shift_map_MF}. For each point in the space, we calculate $\Delta \psi'$ due to $\Delta \mbox{PWV} = 0.5$~mm. We assume a fixed $\alpha=60$~\citep{irwinhilton2005}, a bolometer parameter which is the logarithmic derivative of sensor resistance with respect to temperature, and \fsig $= 8$~Hz, the former value informed by studies of detectors in Advanced ACTPol \citep{2018_Crowley_LTD}. The black contour represents the acceptable upper limit on $\Delta \psi'$ given the assumed parameters. These parameters are taken from Table~\ref{tab:syst_req_table} by identifying the systematic parameter $\Delta \phi_0$ with $\Delta \psi'$. The transparent green band indicates the spread expected for \fdb of the SO detectors. 

In Table~\ref{tab:angle_shift_spec}, we summarize the data from Figure~\ref{fig:angle_shift_map_MF} and those for the other SO detector bandpasses by computing the median value of the angle shifts measured within the green band for all frequencies in the column labeled "$\Delta \psi'\ \mbox{fixed}$''. The spread in this column is the average of the differences between the median and the 16th and 84th percentiles, except for the 225 GHz band where we indicate the spread between the median, and 16th and 84th percentiles, separately. We repeat the acceptable values on $\Delta \phi_0$ from Section~\ref{ssec:results.bias} in the fourth column.

\begin{table}[tb!]
\centering
\begin{tabular}{|c|c|c|c|}
\hline
$\nu_b$ (GHz) & $\Delta \psi' \: \mbox{fixed}$ ($^{\circ}$) & $\Delta \psi' \: \mbox{average}$ spread ($^{\circ}$) & $\Delta \phi_0$ ($^{\circ}$) \\
\hline
27 & $0.03 \pm 0.002$ & $0.008$ & 1.75 \\
40 & $0.01 \pm 0.005$ & $0.002$ & 1.75 \\
94 & $0.04 \pm 0.008$ & $0.01$ & 0.2 \\
148 & $0.09 \pm 0.03$ & $0.03$ & 0.2 \\
225 & $0.2^{+0.25}_{-0.08}$ & $0.06$ & 0.4 \\
280 & $0.1 \pm 0.04$ & $0.03$ & 0.4\\
\hline
\end{tabular}
\caption{Summary statistics of $\Delta \psi'$ given SO detector variation and variation in optical loading from Fig. 17. For each band (left column), we show the median and typical spread of $\Delta \psi'$ values falling within the green band in Fig. \ref{fig:angle_shift_map_MF} (second column). For 225~GHz, we indicate the difference between the median and the 16th and 84th percentiles. These values assume fixed \frot for the HWP, and a fixed typical value for the bolometer parameter $\alpha$. In the third column, we show the spread of $\Delta \psi'$ calculated using 500 draws from the approximate $\Delta$ PWV distribution from 1,000 hours of APEX data are shown in the third column. We collect the angle shift requirements to limit bias on $r$ to 2$\times$10$^{-4}$ from Table~\ref{tab:syst_req_table}.}
\label{tab:angle_shift_spec}
\end{table}

\begin{figure}[tb!]
\includegraphics[width=0.48\textwidth]{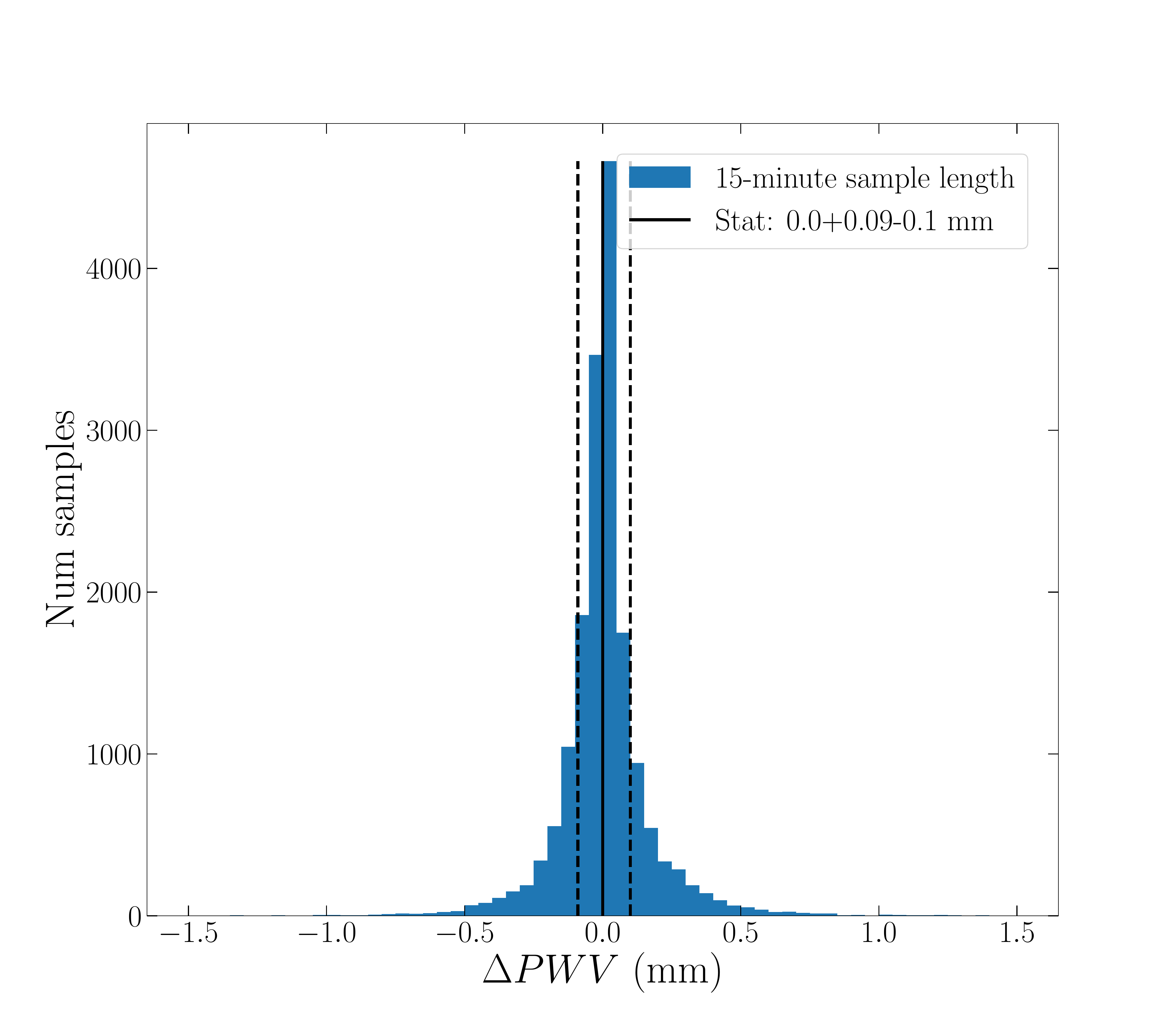}
\includegraphics[width=0.48\textwidth]{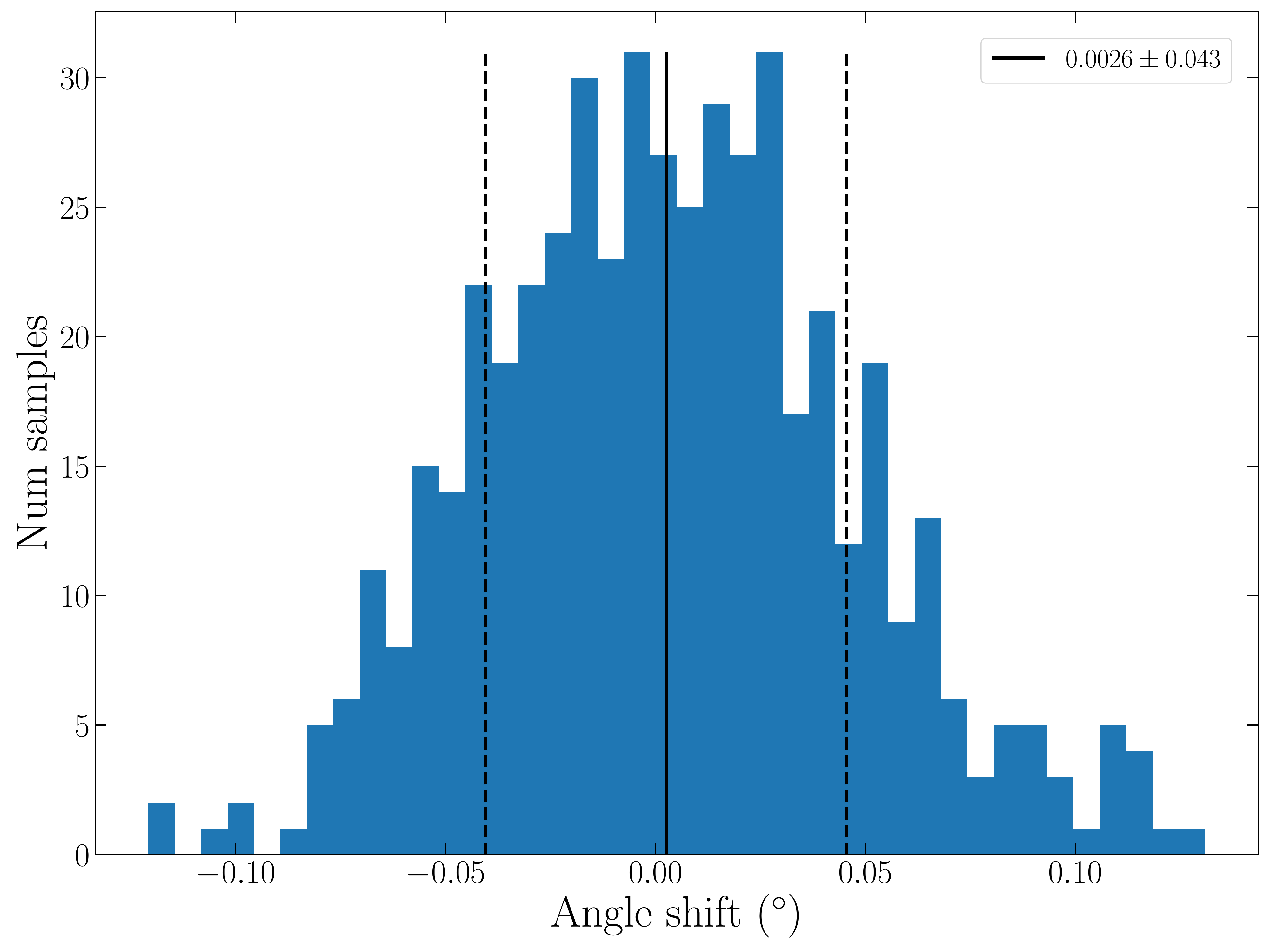}
\caption{Left: Distribution of $\Delta \mbox{PWV}$ at 15-minute intervals over $\sim$4000 hours of APEX data with initial PWV $< 3$~mm, assuming $\Delta \mbox{PWV} =$ 0 at the start of each hour. Right: Recovered $\Delta \psi'$ distribution from 500 draws on a Gaussian model of the $\Delta \mbox{PWV}$ distribution for the SO SAT 225 GHz detector design. The median and spread are reported in the legend. The central value being nearly zero implies minimal bias on $r$ will arise over a season of observing in varying weather conditions.}
\label{fig:ang_shift_avg}
\end{figure}

In the above analysis, we have assumed a fixed and conservatively estimated $\Delta \mbox{PWV}$ value driving $\Delta \psi'$. To account for the averaging effect over observation hours with PWV shifts of different signs, we perform a random draw of 500 PWV drift values. We approximate the distribution of values as normal centered at 0 with $\sigma = 0.1$~mm based on the distribution of the $\Delta \mbox{PWV}$ at 15-minute intervals across the first 1000 hours of APEX data shown in Figure~\ref{fig:ang_shift_avg}. We convert these values to $\Delta \psi'$ assuming fixed detector parameters, those equal to the SO detector design specifications operating at 1.5~mm PWV, and study the central value and spread of the $\Delta \psi'$ directly. We show both the histogram of APEX $\Delta \mbox{PWV}$ values used to determine the approximate Gaussian, and the resulting $\Delta \psi'$ distribution for the SO SAT 225~GHz detector band, in Figure~\ref{fig:ang_shift_avg}. This indicates that the averaging of $\Delta \mbox{PWV}$ seen in Figure~\ref{fig:pwv_data} persists in the $\Delta \psi'$ distribution, which has a median $\Delta \psi' < 3 \times 10^{-3}$ for all bands. The measured widths of these angle shift distributions for each band are also provided in Table~\ref{tab:angle_shift_spec}. This is an additional variance which we do not expect to bias $r$. While this polarization angle jitter may reduce the polarization efficiency of our detectors, it can be calibrated out when SO polarization spectra are cross-correlated with \planck data. Further, the distributions' 95th percentile is within the bounds required for systematic angle errors. We thus conclude that the polarization angle variations due to time constant fluctuations within hour-long observation periods is sufficiently low to neglect and that calibration of the time constants on faster time scales is not necessary. These results indicate that the time-constant calibration period could be increased beyond an hour, but in practice, these measurements are typically performed during the hourly detector biasing.

We also use these studies as an input to define the target detector \fdb specifications. We take a conservative approach and require that the $\Delta \psi'$ for the minimum target value of \fdb does not exceed the requirements in Table~\ref{tab:syst_req_table} up to a PWV of 2.75~mm. The minimum \fdb values from this conservative approach are in Table~\ref{tab:f3db_spec}. We note that SO is using the same detector wafers for both the large-aperture telescope (LAT) and SAT instruments, so the LAT requirements must also be considered in setting the \fdb requirements. The LAT \fdb constraints are set by requiring that the \fdb roll-off of the bolometer matches the Nyquist sampling of the beamwidth as the telescope scans. This gives $f_{\mbox{\scriptsize 3dB, min}}=2.4 \nu_{scan}/\theta_{FWHM}$, where $\nu_{scan}$ is the on-sky telescope scanning frequency ($\sim1$~deg/s) and $\theta_{FWHM}$ is the beam full width at half maximum. We note that the SAT value for \fdb in the 225~GHz band reduces to the LAT value of 144.0~Hz for PWV$<1.75$~mm. It is important to note that both the SAT and LAT inputs push us toward faster detectors than those that have typically been considered for previous CMB experiments, particularly in the UHF bands.

\begin{table}[tb!]
\centering
\begin{tabular}{|c|c|c|}
\hline
$\nu_b$ (GHz) & SAT \fdb (Hz) & LAT \fdb (Hz)  \\
\hline
27 & 5.52 & 19.5  \\
40 & 4.85 & 28.2 \\
94 & 35.0 & 65.5  \\
148 & 104.9 & 102.9  \\
225 & 365.2 & 144.0  \\
280 & 141.0 & 160.0 \\
\hline
\end{tabular}
\caption{SAT and LAT inputs to the \fdb requirements are shown above. The SAT values are set by requiring that the $\Delta \psi'$ for arising from a given detector \fdb does not exceed the requirements in Table~\ref{tab:syst_req_table} at a PWV of 2.75~mm, which is a conservative approach. The LAT requirements are set by the scan rate and beam size and scale linearly with scan speed. For the band centered at 225~GHz, the  SAT value for \fdb reduces to the LAT value of 144.0~Hz for PWV$<1.75$~mm.}
\label{tab:f3db_spec}
\end{table}

\subsection{Calibration strategy}
\label{ssec:cal}
The requirements from Table~\ref{tab:syst_req_table} place calibration requirements on both the polarization angle and bandpass. There are variations in polarization angle and bandpass across detector arrays from the detector fabrication process and telescope optics. These are difficult to reduce, so these quantities must be well-characterized across the arrays. These requirements are significantly more stringent than in previous experiments and in some cases necessitate the development of new calibration technologies and strategies. We note that these requirements also have implications for beam calibration, but these will be discussed in future work.

\subsubsection{Polarization angle calibration}
\label{sssec:pcal}
To characterize the polarization angle of the instruments, SO will use a combination of observations of polarized astronomical sources like Tau~A, artificial sources, and self-calibration. 

Tau~A (the Crab Nebula) emits a polarized synchrotron signal at millimeter wavelengths, which can provide an absolute calibration of the telescope polarization angle. To date, Tau~A has been measured to an uncertainty of $0.33^{\circ}$~\cite{TauA}. \footnote{Note that uncertainties in the calibrator spectrum and any potential frequency or spatial dependence of the angle can add to the uncertainties of the polarization angle measurement. Again, this could be modeled in a similar way to the marginalization scheme above.} Thus, if measured well enough with our instruments, it could meet the requirements for the looser $\Delta r = 10^{-3}$ case. However, for the more stringent case of $\Delta r = 2\times10^{-4}$, this cannot meet the MF requirement of 0.2$^\circ$ alone. 

Given the uncertainty in Tau~A, SO is developing both drone and wire grid calibrators. Artificial polarized sources have achieved uncertainties in polarization angle calibration of $\sim1^{\circ}$~\cite{ABS2018, Takahashi_2008}. 
Mounting a polarized source on a drone could enable polarization angle calibration in the SAT far-field, and integrating these technologies with a star camera for improved position information could open the possibility to meet the $<0.2^{\circ}$ uncertainty requirement~\cite{Nati2017}. We are also developing a wire grid with a gravity reference to provide additional relative and absolute polarization angle measurements at the $<0.2^{\circ}$ level~\cite{ABS2018,Tajima2012}. In the future, CubeSat sources could provide the requisite calibration precision~\cite{johnson2015cubesat}. Since the wire grid calibrator is only present in the near field, we will compare with other calibrators, including Tau-A, as part of the calibration strategy. External calibration will be vital in order to avoid the implicit self-calibration assumption of a null $EB$ signal in the presence of cosmic birefringence.

We also note that we can meet the polarization angle calibration requirement solely through self-calibration~\cite{Keating2013}. Because the polarization angles are a critical calibration, SO will use self-calibration to meet the polarization angle requirements and further verify the polarization angles with Tau~A and artificial sources. This multifaceted approach will ensure that we reach the necessary polarization angle calibration requirements.

\subsubsection{Bandpass calibration}
\label{sssec:bcal}
Roughly once a year, we will take measurements of the spectral response of the detectors in-situ on the telescopes with an FTS. FTS measurements from fielded CMB instruments have demonstrated uncertainties of $<\sim3\%$\footnote{Priv. comm. with Jeff McMahon.}. Section~3.2 shows that the constraints in Table~\ref{tab:syst_req_table} can be relaxed by marginalization, meaning that the current FTS measurement uncertainties would be sufficient (see also \cite{2018ApJ...861...82W}). SO will further improve FTS measurements both through characterizing the FTS instrument response to improve and characterize its systematic performance and developing improved coupling optics that couple the light more cleanly to the telescope optics and ensure that the FTS output fills the detector beams. These improvements will further reduce the uncertainties in FTS measurements.

\section{Conclusions}
\label{sec:conclusion}
The search for primordial $B$-modes is one of the most important pursuits in cosmology. This quest poses both technical and analysis challenges, in the form of stringent requirements on instrument design and calibration, and detailed measurements of Galactic polarized emission.

In this paper, we have studied the impact of instrumental systematic effects on the final constraints on the tensor-to-scalar ratio $r$ for the Simons Observatory. In particular, we have considered the effect of residual systematic effects on the final frequency maps associated with uncertainties in instrument response to polarized, frequency-dependent signals. We have parametrized theses uncertainties in the form of shifts in the mean bandpass frequencies ($\Delta \nu_b$), gain calibration errors ($\Delta g_b$), constant polarization angles ($\Delta\phi_{0,b}$) and linear polarization angle variations as a function of frequency within the band ($\Delta\phi_{1,b}$). We have additionally studied the specific frequency-dependent polarization angles induced by the SO multi-layer HWPs and sinuous antennas.

\begin{table}[tb!]
\centering
\begin{tabular}{| p{0.2\linewidth} | p{0.55\linewidth} | p{0.15\linewidth} |}
\hline
Analysis & Key results & Reference \\
\hline
Biases & MF and UHF bandpasses require sub-percent level calibration, while MF and UHF polarization angle calibration requirements require sub-degree level calibration. LF calibration requirements are several factors looser. & Table~\ref{tab:syst_req_table} \\ \hline
Marginalization & Marginalization scheme eliminates biases without imposing an unacceptable penalty on $\sigma_r$, given previously demonstrated priors on parameters. & Table~\ref{tab:marg} \\ \hline
HWP layers & Bias from 3-layer HWP is within acceptable limits and can be marginalized away. 5-layer HWP not required. & Table~\ref{tab:hwp_spec_angles} \\ \hline
Detector time constants & Polarization angle drifts due to fluctuating optical loading are within the bias requirements given previous optical loading measurements. Requirements on detector $f_{3dB}$ are derived. & Table~\ref{tab:angle_shift_spec}, \ref{tab:f3db_spec} \\ \hline Polarization angle calibration & Requirements met by CMB $EB$ nulling with measurements from Tau~A and artificial polarized sources for redundancy.
 & Section~\ref{sssec:pcal} \\ \hline
Bandpass calibration & Requirements met with bandpass calibration with in-situ FTS measurements and marginalization scheme. & Section~\ref{sssec:bcal} \\ 
\hline
\end{tabular}
\caption{Key findings and results.}
\label{tab:conclusion}
\end{table}

To carry out this study, we have made use of a power-spectrum-based component separation pipeline consisting of $19$ free CMB and foreground parameters describing the signal power spectrum, in addition to the systematic parameters enumerated above.

We have studied the bias on $r$ that these systematics would cause if not included in the signal model to derive conservative constraints to guide the instrument design and calibration. We have determined that the impact of bandpass systematic effects is maximal when they occur in a pair-wise fashion for two nearby frequency bands, and when the systematic shifts take opposite signs in each band, since they are able to mimic variations in foreground spectral indices. In the case of polarization angles, the symmetric shifts in both bands cause the largest bias on $r$, as it produces the same effect as a rotation in the $E$-$B$ plane of a given sky component. In this worst-case scenario, we find that \emph{bandpass uncertainties must be known with sub-percent level accuracy}, which could pose a challenge for FTS-based calibration for future experiments. These requirements could be relaxed significantly for the more likely scenario in which bandpass mis-calibration occurs in a symmetric manner for band pairs due to effects like systematic shifts in the FTS measurements and systematic shifts in the bandpasses from fabrication variation. \emph{Overall polarization angles must be known at the level of a few tenths of a degree}, which could be achievable with current calibration strategies. For SO, the combination of Tau~A, artificial sources, and self-calibration will be used to meet the polarization angle requirements. \emph{The variation of the polarization angle between band edges, on the other hand, must be known to a much lower accuracy of ${\cal O}(10^\circ)$}.

We have found that, although the frequency-dependent polarization angle induced by sinuous antennas and HWPs would cause a bias on $r$ at the level of $\sim10^{-3}$ if not accounted for, this bias is reduced to negligible levels if this frequency dependence is corrected for assuming a CMB spectrum across the full range of frequencies, or if we model it as a linear function within each band.

In most cases, we also observe that it should be possible to detect the presence of these systematics through a simple $\chi^2$ test, which would then motivate modeling them and marginalizing over their uncertainties in the component separation stage of the analysis. This would then remove any bias on $r$ at the cost of potentially increasing its uncertainty. We find, however, that in most cases, assuming reasonable priors on the systematic parameters, the degradation of $\sigma_r$ from marginalization is at the level of $\lesssim$10\%. This is significantly smaller than the degradation expected from additional foreground complexity (at the level of $\sim30\%$ when accounting for foreground frequency decorrelation). \emph{This will allow us to relax the stringent calibration requirements found by our analysis of the induced bias on $r$.}

The SO data will be able to self-calibrate the polarization angle in each frequency by constraining the parity-violating $EB$ correlations. We find that this is the case, even in the presence of foreground $EB$ components, which would have to be at least half as large as the $BB$ amplitude in order to induce a significant bias on $r$.

Our findings have been useful in guiding the design of the SO SATs. We have described the decisions made in the context of 3- versus 5-layered HWPs (where the latter induce an almost frequency-independent polarization angle at the cost of additional mechanical risk) and the impact of detector time constants on the polarization angle uncertainty. These results have also been used to define a calibration strategy for bandpasses and polarization angles in SO.

A number of caveats in our analysis must be noted. Our results are based on a multi-frequency $C_\ell$-based component separation pipeline. This has been used by \bk to derive the current state-of-the-art constraints on $B$-modes, but may suffer from certain shortcomings when applied to data with higher sensitivity over a wider sky patch. The main concern is the problem of spatially-varying foreground spectra, which are difficult to model at the power spectrum level. Although we have accounted for this through the frequency decorrelation parameter, one will need to explore more sophisticated parametrizations, such as a moment expansion~\cite{mangilli2019}. In this scenario, we would expect the qualitative results (i.e.,  the result that the bulk of the additional uncertainty in systematic parameters is absorbed by the foreground model) to hold true, although the overall final uncertainty on $r$ could vary. We have also used a specific parametrization of bandpass uncertainties, through the mean shift and gain degrees of freedom. This is a reasonable description of the expected uncertainties from FTS-based calibrations, but other modes of uncertainty (e.g.,  in the width or slope of the bandpass) could also be relevant. Although the effect of some of these would probably be degenerate with frequency shifts, we have not attempted to  study other parametrizations here. Our work has not studied the impact of uncertainties on the instrumental beam. The beam frequency dependence within the band would couple bandpass uncertainties and the scale dependence of the signal in a non-trivial way that could lead to parameter degeneracies beyond those studied here. We leave the study of this systematic for future work. Our study of polarization angle self-calibration has assumed null CMB $EB$ correlations. We leave the study of simultaneous cosmic birefringence measurement and instrumental polarization angle calibration \citep{2019arXiv190412440M, minamibirefringe} for future work.  Finally, and most importantly, this first study of systematic errors in SO has focused on the impact of instrumental systematics in the form of ``averaged'' uncertainties in the effective global bandpasses and angles of the final set of frequency maps, and thus, we have not modeled the impact of residual temporal or spatial variations in these effects, which could affect some of the conclusions presented here. Studying this will require the use of TOD simulations or potentially faster analytical methods \citep{2020arXiv200800011M}, which we leave for future work.

The increased sensitivities of future experiments, such as CMB Stage-4 \cite{2016arXiv161002743A}, will place additional constraints on the calibration requirements described here, possibly pushing them beyond the limits of current calibration techniques. At this point, modeling the associated systematics and marginalizing over them will likely be necessary in order to achieve reliable constraints on $r$. Our results give us assurance that, in this scenario, the final constraints on $r$ will not be significantly degraded with respect to current forecasts.

\acknowledgments
MHA acknowledges support from the Beecroft Trust and Dennis Sciama Junior Research Fellowship at Wolfson College. This project has received funding from the European Research Council (ERC) under the European Union’s Horizon 2020 research and innovation programme (grant agreement No.~693024).
DA acknowledges support from the Beecroft Trust, and from the Science and Technology Facilities Council through an Ernest Rutherford Fellowship, grant reference ST/P004474/1.
The SISSA group acknowledges support from the ASI-COSMOS network (\url{www.cosmosnet.it}) and the INDARK INFN Initiative (\url{web.infn.it/CSN4/IS/Linea5/InDark}).
EC acknowledges support from the STFC Ernest Rutherford Fellowship ST/M004856/2 and STFC Consolidated Grant ST/S00033X/1, and from the European Research Council (ERC) under the European Union’s Horizon 2020 research and innovation programme (Grant agreement No. 849169).
JC acknowledges support from a SNSF Eccellenza Professorial Fellowship (No. 186879).
YC acknowledges the support from the JSPS KAKENHI Grant Number 18K13558, 18H04347, 19H00674.
This work was supported by the ERC Consolidator Grant {\it CMBSPEC} (No.~725456) as part of the European Union's Horizon 2020 research and innovation program and the Royal Society grant No. URF\textbackslash R\textbackslash 191023.
JE was supported by the French National Research Agency (ANR) grants, ANR-B3DCMB, (ANR-17-CE23-0002), and ANR-BxB (ANR-17-CE31-0022).
HN acknowledges the support from the JSPS KAKENHI Grant Numbers JP17K18785 and JP18H01240.
ZX is supported by the Gordon and Betty Moore Foundation.
This work was supported in part by a grant from the Simons Foundation (Award \#457687, B.K.). This manuscript has been authored by Fermi Research Alliance, LLC under Contract No. DE-AC02-07CH11359 with the U.S. Department of Energy, Office of Science, Office of High Energy Physics.

\bibliography{bibliography}

\end{document}